\documentclass[12pt,reqno]{article}
\usepackage{amsmath}
\usepackage{amsthm}
\usepackage{latexsym}
\usepackage{epsf}
\usepackage{graphicx,color}
\usepackage{amssymb}
\usepackage{epstopdf}
\usepackage{epsfig}

\theoremstyle{remark}
\newtheorem{remark}{Remark}
  
\begin{document}

\newcommand{\A}{{\bf A}}
\newcommand{\B}{{\bf B}}
\newcommand{\bco}{{\boldsymbol{:}}}
\newcommand{\blambda}{{\boldsymbol{\lambda}}}
\newcommand{\bmu}{{\boldsymbol{\mu}}}
\newcommand{\bn}{{\bf n}} 
\newcommand{\bs}{{\bf s}}
\newcommand{\bnabla}{{\boldsymbol{\nabla}}}
\newcommand{\bomega}{{\boldsymbol{\omega}}}
\newcommand{\bsigma}{{\boldsymbol{\sigma}}}
\newcommand{\btheta}{{\boldsymbol{\theta}}}
\newcommand{\btau}{{\boldsymbol{\tau}}}
\newcommand{\bxi}{{\boldsymbol{\xi}}}

\newcommand{\bo}{{\overline{B(0)}}}
\newcommand{\bu}{{\bf u}}
\newcommand{\bU}{{\bf U}}
\newcommand{\bv}{{\bf v}}
\newcommand{\bw}{{\bf w}}
\newcommand{\bzero}{{\bf 0}}
\newcommand{\ct}{{\mathcal{T}}}
\newcommand{\cth}{{\mathcal{T}_h}}
\newcommand{\dsum}{{\displaystyle\sum}}
\newcommand{\bC}{{\bf C}}
\newcommand{\D}{{\bf D}}
\newcommand{\e}{{\bf e}}
\newcommand{\F}{{\bf F}}
\newcommand{\G}{{\bf G}}
\newcommand{\g}{{\bf g}}

\newcommand{\Gx}{{{\overrightarrow{\bf Gx}}}}
\newcommand{\Gtx}{{{\overrightarrow{\bf G}(t){\bf x}}}}

\newcommand{\I}{{\bf I}}
\newcommand{\intbt}{{\displaystyle{\int_{B(t)}}}}

\newcommand{\intG}{{\displaystyle{\int_{\Gamma}}}}
\newcommand{\into}{{\displaystyle{\int_{\Omega}}}}
\newcommand{\intobt}{{\displaystyle{\int_{\Omega\setminus\overline{B(t)}}}}}
\newcommand{\intpb}{{\displaystyle{\int_{\partial B}}}}
\newcommand{\lto}{{L^2(\Omega)}}
\newcommand{\no}{{\noindent}}
\newcommand{\obo}{{\Omega \backslash \overline{B(0)}}}
\newcommand{\obt}{{\Omega \backslash \overline{B(t)}}}
\newcommand{\oo}{{\overline{\Omega}}}
\newcommand{\R}{{\text{I\!R}}}
\newcommand{\T}{{\bf T}}
\newcommand{\V}{{\bf V}}
\newcommand{\w}{{\bf w}}
\newcommand{\W}{{\bf W}}
\newcommand{\x}{{\bf x}}
\newcommand{\Y}{{\bf Y}}
\newcommand{\y}{{\bf y}}
\newcommand{\fr}{\color{red}}

\parskip 4pt
\abovedisplayskip 7pt
\belowdisplayskip 7pt
\parindent=24pt

\noindent{\Large\bf  A 3D DLM/FD method for simulating the  motion of spheres in a
bounded shear flow of Oldroyd-B fluids}

\bigskip
\normalsize \noindent{Shang-Huan Chiu$^a$,  Tsorng-Whay Pan$^{a,}$\footnote{Corresponding author: e-mail:  pan@math.uh.edu, tel.: 713-743-3448},  
 Roland Glowinski$^{a,b}$} 
\vskip 2ex
\noindent{\small $^a$ Department of Mathematics, University of Houston, Houston,
Texas  77204, USA}  
\vskip 0.25ex
\noindent{\small $^b$ Department of Mathematics,  Hong Kong Baptist University, Kowloon Tong, Hong Kong}  
\vskip 2ex
\noindent {\bf Abstract}
\vskip 1ex

We present a novel distributed Lagrange multiplier/fictitious domain (DLM/FD)
method for simulating fluid-particle interaction in Oldroyd-B fluids under
creeping conditions.  The results concerning two ball interaction in a three dimensional (3D) 
bounded shear flow are obtained for  {Weissenberg numbers}  up to 1 . The pass and return 
trajectories of the two ball mass centers are similar to those in a Newtonian fluid; but 
they lose the symmetry due to the effect of elastic force arising from viscoelastic fluids. 
A  tumbling chain of two balls {(a dipole)}  may occur, depending on the value of the  Weissenberg 
number and the initial vertical displacement of the ball mass center to the middle plane
between two walls. 


\section{Introduction}

Particles suspended in flowing fluids occur in many engineering and biological systems. 
The rheological behavior of suspensions  has been studied heavily during recent decades (e.g., see
\cite{Denn2014}   for the overview of  the rheology of suspensions).  For the dynamics of 
rigid non-Brownian particles suspended in viscoelastic fluids, D'Avino and Maffettone have  
reviewed the existing literature in \cite{DAvino2015} with focus on theoretical predictions, 
experimental observations and numerical simulations of peculiar phenomena induced by fluid 
elasticity which dramatically {affects} the particle motion and patterning. In Newtonian fluids, 
random displacements resulting from particle encounters under creeping conditions lead to 
hydrodynamically induced particle migration, which constitutes an important mechanism for particle 
redistribution in the suspending fluid (see, e.g., \cite{Zurita2007} and the references therein).  
But particle suspensions  in viscoelastic fluids have different behaviors, e.g., strings 
of spherical particles aligned in the flow direction (e.g., see \cite{Michele1977,  Scirocco2004, Won2004,  
Pasquino2013}) and   2D crystalline patches of particles along the flow direction \cite{Pasquino2010} in shear flow. 
As mentioned in \cite{Snijkers2013},  these flow-induced self-assembly phenomena have great potency for 
creating ordered macroscopic structures  by exploiting  the complex rheological properties of the suspending 
fluid as driving forces, such as its shear-thinning and elasticity. To understand more about particle interaction 
in viscoelastic fluids,  Snijkers {\it et al.}   \cite{Snijkers2013} have studied experimentally the two ball 
interaction  in Couette flow of viscoelastic fluids in order to understand flow-induced assembly behavior associated 
with the string formation. They obtained that, in high-elasticity Boger fluid, the pass trajectories have a zero radial shift, but are not
completely symmetric. In a wormlike micellar surfactant with a single dominant relaxation time and a broad spectrum 
shear-thinning elastic polymer solution, interactions are highly asymmetric and both pass and return trajectories have 
been obtained.      Based on their observation, shear-thinning of the viscosity seems to be the key rheological 
parameter that determines the overall nature of the hydrodynamical interactions, rather than the relative magnitude of the normal 
stress differences.  Same conclusion about the role of shear-thinning on  the aggregation of many particles has been reported
in \cite{Scirocco2004, Won2004}.  There are numerical studies of the two particle interaction and aggregation in 
viscoelastic fluids. {For example,} Hwang {\it et al.} \cite{Hwang2004} applied a finite element {method} to perform a two-dimensional (2D) 
computational study and obtained the existence of complex kissing-tumbling-tumbling interactions for two 
disks in an Oldroyd-B fluid in sliding bi-periodic frames: The two circular disks keep rotating around each other 
while their {centers} come closer to each other.  Choi {\it et al.} \cite{Choi2010} used an extended finite 
element method {(a methodology introduced in Mo\"es {\it et al.} \cite{Moes1999})} to simulate two  circular particles in a 2D bounded shear 
flow between two moving walls for a Giesekus fluid:
Besides {the fact} that the two disks either pass each other, have reversing trajectories (return) or rotate as a pair (tumble), they also had 
another interaction, {namely,} the two disks {rotate} at a constant speed with their mass centers {remaining} at a fixed position.  To simulate the 
interaction of two spherical particles interacting in an Oldroyd-B fluid, Yoon {\it et al.} \cite{Yoon2012} applied a finite element 
method to discretize {the} fluid flow with a discontinuous Galerkin approximation for {the} polymer stress: In their numerical approach, the rigid 
property of the particles is imposed by treating them as a fluid having a much higher viscosity than the surrounding fluid. They obtained that, 
for the two balls initially located in the same vorticity plane, the balls either pass, return, or tumble in a bounded shear flow driven by 
two moving walls for the  Weissenberg number up to 0.3.  To study numerically the alignment of two and three balls in a viscoelastic fluid,
Jaensson {\it et al.} \cite{Jaensson2016} developed a computational method which mainly combines the finite element method, the arbitrary 
Lagrange-Euler method \cite{Hu2001}, the log-conformation representation for the conformation tensor \cite{Fattal2004, Hulsen2005}, 
SUPG stabilization  \cite{Brooks1982} and second-order time integration schemes. Using such computational method, they simulated the motion of two and three 
balls in bounded shear flows of a viscoelastic fluid of Giesekus type with the effect of the shear-thinning. They concluded 
that the presence of normal stress differences is essential for particle alignment to occur, although it is strongly promoted 
by shear thinning. 

To simulate  the interaction of neutrally buoyant balls in a bounded shear flow of Oldroyd-B fluids in three  dimensions (3D), we have  generalized a DLM/FD 
method developed in  \cite{Pan2015} for simulating the motion of neutrally buoyant particles in Stokes flows of Newtonian fluids to 3D and then combined such 
method with {an} operator splitting scheme and {a} matrix-factorization approach for treating numerically the constitutive equations of the conformation 
tensor of Oldroyd-B fluids. In this matrix-factorization approach \cite{Hao2009}, which is {a} technique closely related to the one developed by Lozinski 
and Owens in \cite{lozinski2003}, we solve the equivalent equations for the conformation tensor so that the positive definiteness of the conformation tensor at
the discrete time level can be preserved. This aforementioned method  has been validated by comparing the numerical results 
of the ball rotating velocity in shear flow with the available results in literature. For the encounter of two balls in a bounded shear flow, the trajectories 
of the  two ball mass centers are consistent with those obtained in \cite{Yoon2012}. We have further tested the cases of two ball interaction for the Weissenberg 
number up to 1; {our results show the two balls either passing, returning, or tumbling} in a bounded shear flow driven by two moving walls. The passing 
over/under trajectories of the two ball mass centers {loses its} symmetry due to the effect of {the} elastic force arising from Oldroyd-B fluids. 
While two balls form a chain and then tumble  in a shear flow driven by two walls, the tumbling motion can change to kayaking for higher Wi.
The content of the article is as follows: We discuss the DLM/FD formulation 
and then the related numerical schemes in Section 2. In Section 3, we first validate our methodology by comparing numerical results 
{for} particle motion  with those available in literature. We also present the results of numerical 
simulations investigating the interaction of two balls in a bounded shear flow. Conclusions are summarized in Section 4.

\section{Models and numerical methods}

\subsection{DLM/FD formulation}
Fictitious domain formulations using distributed Lagrange multiplier for flow around freely moving
particles at finite Reynolds numbers and their associated computational methods have been developed and tested in, e.g., 
\cite{RG1999, RG2000, RG2001, RG2003, Pan2002, Pan2005, Pan2008}.
For the cases of a neutrally buoyant particle in two-dimensional fluid flows of a Newtonian fluid at the Stokes regime, a 
similar DLM/FD method has been developed and validated in \cite{Pan2015}. In this section, we  discuss first 
the formulation for the case of a ball and then the associated  numerical treatments for simulating its motion
in a 3D bounded  shear flow of Oldroyd-B fluids.  Let $\Omega \subset \R^3$ be a rectangular parallelepiped  filled with an Oldroyd-B 
fluid and containing a freely moving rigid sphere $B$ centered at $\G=\{G_1, G_2, G_3\}^t$. 
\begin{figure}
\begin{center}
\leavevmode
\epsfxsize=3.25in
\epsffile{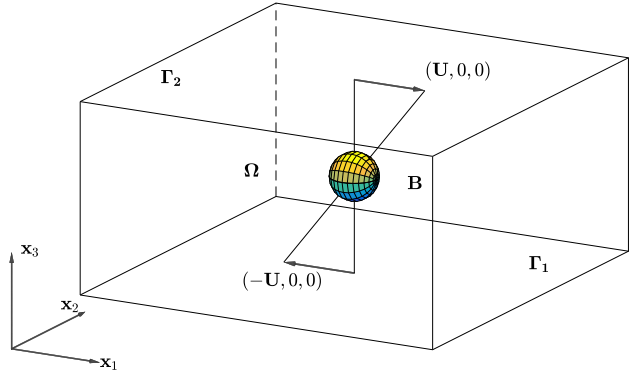}
\end{center}
\caption{An example of a shear flow region with one {ball}.}\label{fig:1}
\end{figure}
\noindent The governing equations  are presented in the following 
\begin{eqnarray}
&& - \bnabla \cdot \bsigma^s - \bnabla \cdot \btau ={\bf g} \ \ in \ \ \Omega\setminus \overline{B(t)},
\, \, t \in (0,T),
\label{eqn:1}\\
&& {\boldsymbol{\nabla}} \cdot {\bf u}=0 \ \ in \ \ \Omega\setminus \overline{B(t)}, \, \,
t \in (0,T),
\label{eqn:2}\\
&& {\bf u}= {\bf g}_0 \ \ on \ \ \Gamma \times (0,T), \ with \ \displaystyle\int_\Gamma {\bf g}_0 \cdot {\bf n} \,d \Gamma =0, \label{eqn:3}\\
&& \dfrac{\partial \bC}{\partial t} + ( \bu \cdot \bnabla ) \bC  -  (\bnabla \bu) \bC -  \bC (\bnabla \bu)^t    =  -  \dfrac{1}{\lambda_1} (\bC-\I) 
\ \text{in}\  \Omega \setminus \overline {B(t)},  \label{eqn:4}  \\
&& \bC(\x,0)=\bC_0(\x), \ \x \in \Omega \setminus \overline {B(0)},  \ \
\bC=\bC_L \ on \ \Gamma^-.   \label{eqn:5} 
\end{eqnarray}
In (\ref{eqn:1}), $\g$ denotes gravity and  the Cauchy stress tensor $\bsigma$ is splitted into two parts, a Newtonian (solvent) part  $\bsigma^s$ and 
a viscoelastic part $\btau$, {with}:
\begin{eqnarray*}
 &&\bsigma^s=-p \I + 2 \mu {\bf D}(\bu), \\ 
 &&\btau=\dfrac{\eta}{\lambda_1}  (\bC-\I),
\end{eqnarray*}
where ${\bf D}(\bu)= (\bnabla\bu + (\bnabla\bu)^t)/2$ is the rate of deformation tensor, $\bu$ is the flow velocity, $p$ is the pressure,  $\bC$  is the conformation tensor, 
$\I$ is the identity tensor, $\mu=\eta_1\lambda_2/\lambda_1$ is the solvent viscosity of the fluid, $\eta=\eta_1-\mu$ is the elastic viscosity of the fluid, 
$\eta_1$ is the fluid viscosity,  $\lambda_1$ is the relaxation time of the fluid, and $\lambda_2$ is the retardation time of the fluid.
The conformation tensor $\bC$ is symmetric and positive definite (see, e.g., \cite{joseph1990}).
In (\ref{eqn:3}), $\Gamma$  is the union of the bottom boundary $\Gamma_1$ and top boundary   $\Gamma_2$ as in 
Figure \ref{fig:1} and $\bn$ is the unit normal vector pointing outward to the flow region,
$\Gamma^{-}(t)$ in (\ref{eqn:5}) being the upstream portion of $\Gamma$ at time $t$.  
The boundary conditions given in (\ref{eqn:3}) are ${\bf g}_0=\{-U, 0, 0\}^t$ on $\Gamma_1$ and ${\bf g}_0=\{U, 0, 0\}^t$ on $\Gamma_2$ for a
bounded shear flow.  We assume also that the flow is periodic in the $x_1$ and $x_2$ directions with 
the periods $L_1$ and $L_2$, respectively,  a no-slip condition taking place on the boundary of the particle 
$\gamma(=\partial B)$, namely
\begin{equation}
{\bf u}({\bf x}, t)={\bf V}(t) +  {\bomega}(t) \times \overrightarrow{{\bf G}(t){\bf x}}, \ \forall \ {\bf x} \in \partial {B(t)}, 
\, t \in (0,T) \label{eqn:6}
\end{equation}
with $\overrightarrow{{\bf G}(t){\bf x}} =\{x_1-G_1(t),x_2-G_2(t),x_3-G_3(t)\}^t$. In addition to (\ref{eqn:6}), the motion of particle
 $B$ satisfies the following Euler-Newton's equations
\begin{eqnarray}
&& \dfrac{d {\bf G}}{d t}={\bf V}, \label{eqn:7} \\
&& \dfrac{d \btheta }{d t}={\bomega}, \label{eqn:8} \\
&& M_p \dfrac{d \bf V}{d t} = M_p \, {\bf g}+{\bf F}_H, \label{eqn:9} \\
&& {\bf I}_p \dfrac{d {\bomega} }{dt} =\T_H, \label{eqn:10} \\
&&  {\bf V}(0)={\bf V}_0, \ {\bomega}(0) = {\bomega}_0,{\bf G}(0)={\bf G}_0, \ \btheta(0)=\btheta_0, \label{eqn:11}
\end{eqnarray}
where $M_p$ and  ${\bf I}_p$ are the mass and inertia tensor of $B$, respectively, 
$\bf V$ is the velocity of the center of mass, ${\bomega}$
is the angular velocity and $\btheta$ is the inclination angle of the particle. 
The hydrodynamical forces and torque are given by
\begin{equation}
{\bf F}_H=-\displaystyle\int_{\partial B} \boldsymbol{\sigma} {\bf n}\, ds, \ \ 
\T_H=-\displaystyle\int_{\partial B} \Gx \times \boldsymbol{\sigma} {\bf n} \,ds.
\label{eqn:12}
\end{equation}

\noindent To obtain a distributed Lagrange multiplier/fictitious domain formulation
for the above problem (\ref{eqn:1})--(\ref{eqn:12}), we proceed as in
\cite{RG1999, RG2001}, namely: (i) we derive first a global variational formulation (of the virtual power type) of problem 
(\ref{eqn:1})--(\ref{eqn:12}), (ii) we then fill the region occupied by the rigid body 
by the surrounding fluid (i.e., embed $\Omega\setminus\overline{B(t)}$ in $\Omega$) with
the constraint that the fluid inside the rigid body region has a rigid 
body motion, and then (iii) we relax the rigid body motion constraint by using a distributed 
Lagrange multiplier, obtaining  thus  the following fictitious domain formulation over the entire region $\Omega$:
 
\vskip 2ex
 
\noindent For  a.e.  $t \in (0,T)$,  find
$\bu(t) \in \V_{{\bf g}_0}$, $p(t) \in L_0^2 (\Omega)$,  $\bC(t)\in \V_{\bC_{L}(t)}$  $\V(t) \in
\R^3$,  $\G(t) \in \R^3$,  $\bomega(t) \in \R^3$,  $\blambda(t) \in
\Lambda(t)$ such that 
\begin{eqnarray}
&&\begin{cases}
- \into p \bnabla \cdot \bv d\x + 2 \mu \into  {\bf D}(\bu) \boldsymbol{:} {\bf D}(\bv) \, d\x - \into  (\bnabla \cdot \btau)\cdot \bv   \, d\x \\
\hskip 20pt  -< \blambda, \bv - \Y - \bxi \times  \Gx >_{\Lambda(t)} +  M_p \dfrac{d\V}{dt} \cdot
\Y +  {\bf I}_p  \dfrac{d \bomega}{dt}  \cdot \bxi \\
=(1 - \dfrac{\rho_f}{\rho_s}) M_p \ \g \cdot \Y + \rho_f \into \g \cdot \bv d\x,  \ \forall \bv \in {\bf V}_0, \ \
\forall \Y \in \R^3, \ \ \forall \bxi \in \R^3,\end{cases} \label{eqn:13}
\end{eqnarray}
\begin{eqnarray}
&&\into q \bnabla \cdot \bu(t) d\x = 0, \ \forall q \in L^2(\Omega),\label{eqn:14} \\
&&<\bmu, \bu(t) - \V(t) - \bomega(t) \times  \Gx >_{\Lambda(t)} =0, \ \forall \bmu \in \Lambda(t),\label{eqn:15} \\
&& \int_{\Omega} \left( \dfrac{\partial \bC }{\partial t} + (\bu \cdot \bnabla)\bC   - (\bnabla \bu)\bC 
- \bC (\bnabla \bu)^t \right)  : \bs\, d\x \label{eqn:16}\\ 
&&\hskip 20pt =  - \int_{\Omega}  \dfrac{1}{\lambda_{1}}(\bC- \I)  : \bs\,  d\x, \forall \bs \in \V_{\bC_0}, 
\text{with}\ \bC=\I \ \text{in} \ B(t), \nonumber \\
&&\dfrac{d\G}{dt}=\V,  \label{eqn:17} \\
&&\bC(\x, 0) = \bC_0(\x), \forall \x \in \Omega, , \text{with}\ \bC_0=\I \ \text{in} \ B(0),\label{eqn:19}\\ 
&&\G(0) = \G_0, \ \V(0) = \V_0, \ \bomega(0) = \bomega_0, \ B(0) = B_0, \label{eqn:20}
\end{eqnarray}
where the function spaces in problem (\ref{eqn:13})--(\ref{eqn:20}) are defined by
\begin{eqnarray*}
&&\V_{{\bf g}_0} = \{\bv|\bv \in (H^1(\Omega))^3, \ \bv = {\bf g}_0 \ on \ \Gamma, \text{ $\bv$ is periodic in the $x_1$ and $x_2$ } \\
&& \hskip 0.65in  \text{ directions with periods $L_1$ and $L_2$, respectively}\}, \\
&&\V_0 = \{\bv|\bv \in (H^1(\Omega))^3, \ \bv = {\bf 0} \ on \ \Gamma, \text{ $\bv$ is periodic in the $x_1$ and $x_2$ } \\
&& \hskip 0.65in \ \text{ directions with periods $L_1$ and $L_2$, respectively}\}, \\
&&L_0^2 (\Omega) = \{ q|q \in L^2 (\Omega), \ \int_{\Omega} q \,d\x = 0\}, \\
&& \V_{\bC_{L}(t)}=\{ \ \bC\ |\ \bC\ \in ({H^1(\Omega)})^{3\times 3},  \bC=\bC_{L}(t) \ \text{on}\ \Gamma^{-}\},  \\
&& \V_{\bC_{0}}=\{ \ \bC\ |\ \bC\ \in ({H^1(\Omega)})^{3\times 3},  \bC={\bf 0} \ \text{on}\ \Gamma^{-}\},  \\
&&\Lambda(t)= (H^1(B(t)))^3,
\end{eqnarray*}
and for any $\bmu  \in  H^1(B(t))^3$ and any $\bv \in {\bf V}_0$, the  pairing
$<\cdot,\cdot>_{\Lambda(t)}$ in (\ref{eqn:13}) and (\ref{eqn:15})  is defined by
$$
 <\bmu, \bv>_{\Lambda(t)} = \displaystyle\int_{B(t)} (\bmu \cdot \bv
 + d^2 \bnabla \bmu : \bnabla  \bv) \,d\x
$$
where $d$ is a scaling constant, a typical choice for $d$ being the diameter of particle $B$.  
\begin{remark}
In relation (\ref{eqn:13})  we can replace
$2\into {\bf D}(\bu) {\boldsymbol :} {\bf D} (\bv)\, d\x$
by $\into \bnabla \bu  {\boldsymbol :} \bnabla \bv\, d\x$.
Also the gravity term $\g$ in (\ref{eqn:13}) can be absorbed into
the pressure term.  
\end{remark}

\begin{remark}
In the system (\ref{eqn:13})--(\ref{eqn:20}), the treatment of neutrally buoyant particles is quite different from those considered 
in, e.g., \cite{Pan2002, Pan2005} for the cases of neutrally buoyant particles in incompressible 
viscous flow  modeled by the full Navier-Stokes equations.  For the particle-flow
interaction under creeping flow  conditions considered in this article, there is no need to add any extra constraint on the 
Lagrange multiplier as in \cite{Pan2002, Pan2005}.
\end{remark}
 
\subsection{\bf Numerical methods}
For the space discretization, we have chosen $P_1$-$iso$-$P_2$ and $P_1$ finite element spaces for 
the velocity field and pressure, respectively, (like in Bristeau et al.
\cite{Bristeau1987} and Glowinski \cite{RG2003}), that is
\begin{eqnarray*}
&&\W_h=\{\bv_h| \bv_h \in (C^0(\overline{\Omega}))^3, \
\bv_h|_T \in (P_1)^3, \  \forall T \in \cth,  \ \bv_h \ \text{\it is periodic in the $x_1$ }\\
&&\hskip 0.5in  \text{\it and $x_2$ directions with the periods $L_1$ and $L_2$, respectively }  \},\\
&&\W_{0h}=\{\bv_h| \bv_h \in \W_h, \ \bv_h={\bf 0} \ on \ \Gamma\},\\
&&L^2_h=\{q_h|q_h\in C^0(\overline{\Omega}), \ q_h|_T\in P_1,
\ \forall T\in \ct_{2h}, \ q_h  \ \text{\it is periodic in the $x_1$ }\\
&&\hskip 0.5in  \text{\it and $x_2$ directions with the periods $L_1$ and $L_2$, respectively } \},\\
&&\displaystyle L^2_{0h}=\{q_h|q_h\in L^2_h,\ \int_{\Omega} q_h\,d\x =0\},
\end{eqnarray*}
where $h$ is the space {discretization} mesh size, $\cth$ is a regular tetrahedral mesh {covering} $\Omega$, $\ct_{2h}$ is another tetrahedral 
mesh {also covering} $\Omega$, twice coarser than $\cth$, and $P_1$ is the space of the polynomials in three variables of degree $\le 1$.

The  finite dimensional spaces  for approximating  $\V_{\bC_{L}(t)}$ and  $\V_{\bC_{0}}$, respectively, are defined by
 \vskip -0.25in
\begin{eqnarray*}
&& \hskip -20pt \V_{\scriptscriptstyle \bC_{L_h}(t)} = \{  \bs_h | \bs_h \in  ({C^0(\overline{\Omega})})^{\scriptscriptstyle 3\times 3}, 
\bs_h |_{\scriptscriptstyle T} \in  (P_1)^{\scriptscriptstyle 3\times 3}, \forall T  \in  \cth,  \bs_h|_{\scriptscriptstyle \Gamma_h^{-}} = \bC_{L_h}(t),\ \bs_h  \ \text{\it is periodic} \\
&&  \hskip 0.5in     \text{\it in the $x_1$   and $x_2$ directions with the periods $L_1$ and $L_2$, respectively } \},  \\
&& \hskip -20pt\V_{\scriptscriptstyle \bC_{0h}} = \{ \bs_h | \bs_h \in  ({C^0(\overline{\Omega})})^{\scriptscriptstyle 3\times 3}, 
\bs_h |_{\scriptscriptstyle T} \in  (P_1)^{\scriptscriptstyle 3\times 3}, \forall T
\in  \cth,  \bs_h|_{\Gamma_h^{-}}={\bf 0}\ \bs_h  \ \text{\it is periodic}  \\
&&  \hskip 0.4in    \text{\it  in the $x_1$   and $x_2$ directions with the periods $L_1$ and $L_2$, respectively }  \},
\end{eqnarray*}
where  $\Gamma_h^{-}=\{\x \ |\ \x \in \Gamma,\ \g_{0h}(\x)\cdot \bn(\x)<0\}$. 
\begin{figure}[t!]
\begin{center}
\leavevmode
\epsfxsize=1.75in
\epsffile{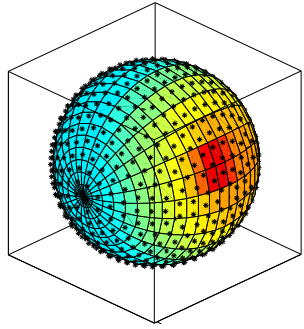}
\end{center}
\caption{An example of collocation points chosen {on}  $\partial B$.}\label{fig:2}
\end{figure}
\vskip 2ex

For simulating the particle motion in fluid flows, a typical finite dimensional space approximating $\Lambda(t)$ 
(e.g., see \cite{RG2001, Pan2005, Pan2008}) is defined as follows:  let $\{\y_i\}_{i=1}^{N(t)}$ be a set of points
covering $\overline{B(t)}$; the discrete multiplier space $\Lambda_{h}(t)$ 
is defined by
\begin{equation}
\Lambda_{h}(t)=\{\bmu_h|\bmu_h=\sum\limits_{i=1}^{N(t)}\bmu_i
\delta(\x-\y_i), \ \bmu_i\in\R^3, \ \forall i=1,...,N(t)\},
\label{eqn:21}
\end{equation}
where $\delta(\cdot)$ is the Dirac measure at $\x={\bf 0}$.
Then, we define a pairing over  $\Lambda_{h}(t) \times  \W_{0h}$ by
\begin{equation}
<\bmu_h,\bv_h>_{\Lambda_h(t)} = \sum\limits_{i=1}^{N(t)}
\bmu_i\cdot\bv_h(\y_i), \ \forall \bmu_h \in \Lambda_{h}(t),
\ \bv_h\in \W_{0h}.
\label{eqn:22}
\end{equation}
A typical set $\{\y_j\}_{j=1}^{N(t)}$ of the points of $\overline{B}(t)$ to be used in  (\ref{eqn:22}) is defined as
\begin{equation*}
 \{\y_j\}_{j=1}^{N(t)}=\{\y_j\}_{j=1}^{N_1(t)} \cup \{\y_j\}_{j=N_1(t)+1}^{N(t)}
\end{equation*}
where $\{\y_j\}_{j=1}^{N_1(t)}$ (resp., $\{\y_j\}_{j=N_1(t)+1}^{N(t)}$)  is the set of those vertices of the velocity grid  $\cth$  contained in
$B(t)$ and whose distance to $\partial B(t) \ge h/2$  (resp., is a set of selected points of $\partial B(t)$, as shown in
Fig. \ref{fig:2}). But for simulating   particle interactions in Stokes flow, we have modified the discrete pairing $<\cdot,\cdot>_{\Lambda_h(t)}$ 
as follows:
\begin{equation}
<\bmu_h,\bv_h>_{\Lambda_h(t)} 
= \sum\limits_{i=1}^{N_1(t)} \bmu_i\cdot\bv_h(\y_i) + \sum\limits_{i=N_1(t)+1}^{N(t)}\sum\limits_{j=1}^M  \ \bmu_i \cdot \bv_h(\y_i) \ D_h(\y_i-\x_j)\ h^3, 
\label{eqn:23}
\end{equation}
for  $\bmu_h \in \Lambda_{h}(t)$ and $\bv_h\in \W_{0h}$ where $h$ is the uniform finite element mesh size 
for the velocity field, $\{\x_j\}_{j=1}^M$ is the set of the grid points of the velocity field,  and 
the function $D_{h}({\bf X}-\bxi)$ is defined as
\begin{equation}
D_{h}({\bf X}-{\bxi})=\delta_{h}(X_{1}-\bxi_{1})\delta_{h}(X_{2}-\bxi_{2})\delta_{h}(X_{3}-\bxi_{3})\label{eqn:24}
\end{equation}
with ${\bf X}=\{X_1,X_2,X_3\}^t$,  ${\bxi}=\{\bxi_1,\bxi_2,\bxi_3\}^t$,   the one--dimensional  approximate Dirac measure $\delta_h$  being defined by 
\begin{equation}
\delta_{h}(s)=\begin{cases}
\frac{1}{8h}\left(3-\frac{2|s|}{h}+\sqrt{1+\frac{4|s|}{h}-4(\frac{|s|}{h})^2}\right), &   \ \lvert s \rvert \le h, \\
\frac{1}{8h}\left(5-\frac{2|s|}{h}-\sqrt{-7+\frac{12|s|}{h}-4(\frac{|s|}{h})^2}\right), &  \ h \le \ \lvert s \rvert \le 2h, \\
0, & \  \text{otherwise}.
\end{cases}\label{eqn:25}
\end{equation}
The above approximate delta functions $\delta_h$ and $D_h$ are the typical ones used in the popular immersed boundary method developed 
by Peskin, e.g, \cite{Peskin1977, Peskin1980, Peskin2002}.

To fully discretize system (\ref{eqn:13})--(\ref{eqn:20}), we reduce it first to a finite dimensional initial value problem using the above finite 
element spaces (after dropping most of the sub-scripts $h$'s). Next, we combine the Lozinski-Owens factorization approach 
(see, e.g., \cite{lozinski2003}, \cite{Hao2009})  with the Lie scheme (e.g., see \cite{Chorin1978}, \cite{Tai2016}, and \cite{Osher2016}) to decouple 
the above finite element analogue of system  (\ref{eqn:13})--(\ref{eqn:20}) into a sequence of subproblems and apply the backward Euler schemes to 
time-discretize some of these subproblems. Finally we obtain thus the following sequence of sub-problems (where $\triangle t (> 0)$ is a 
time-discretization step and $t^n = n\triangle t$):
 
\begin{equation} 
\bC^{0}={\bC}_{0}, {\bf G}^0={\bf G}_0,  {\bf V}^0={\bf V}_0, \text{and} \ {\bomega}^0={\bomega}_0\  \text{are given}; 
\label{eqn:26}
\end{equation}

\noindent For $n \ge 0$,  $\bC^{n}$, ${\bf G}^n$, ${\bf V}^n$, ${\bomega}^n$  being known, 
we compute the approximate solution at $t=t^{n+1}$ via the following fractional steps:

\begin{enumerate}

\item {We first predict the position and the translation velocity of the center of mass as follows: }
\begin{eqnarray}
&&  \dfrac{d {\bf G}}{d t} = {\bf V}(t),\label{eqn:30}\\
&& M_p\dfrac{d {\bf V}}{d t} = {\bf 0},\label{eqn:31}\\
&& \I_p \dfrac{d  \bomega}{d t}={\bf 0},\label{eqn:32}\\
&& {\bf V}(t^n)={\bf V}^n, \bomega(t^n)=\bomega^{n}, {\bf G}(t^n)={\bf G}^n, 
\label{eqn:33} 
\end{eqnarray}
\noindent for $t^n < t < t^{n+1}$.  Then set ${\bf V}^{n+\frac{1}{4}}={\bf V}(t^{n+1})$, 
$ \bomega^{n+\frac{1}{4}}=\bomega(t^{n+1})$, and ${\bf G}^{n+\frac14}={\bf G}(t^{n+1})$. 
After the center ${\bf G}^{n+\frac14}$ is known, the position $B^{n+\frac14}$ occupied by the particle is {determined}.

\item {Next,  we enforce the rigid body motion in $B^{n+\frac14}$ and solve for ${\bf u}^{n+\frac24}$, $p^{n+\frac24}$, ${\bf V}^{n+\frac24}$ and $\bomega^{n+\frac24}$ 
simultaneously as follows:}

Find $\bu^{n+\frac24}\in \W_h$, $\bu^{n+\frac24}=\g_{0h}$  on $\Gamma$, 
$p^{n+\frac24}\in L^2_{0h}$, $\blambda^{n+\frac24}\in \Lambda_{h}^{n+\frac14}$, 
${\bf V}^{n+\frac24}\in\R^3$, $\bomega^{n+\frac24}\in\R^3$ so that 
\begin{eqnarray}
&&\hskip -12pt \begin{cases}
-\into p^{n+\frac24} \bnabla \cdot\bv d\x   + \mu \into \bnabla\bu^{n+\frac24}:\bnabla \bv\, d\x - \into  (\bnabla \cdot \dfrac{\eta}{\lambda_1}  (\bC^{n}-\I))\cdot \bv   \, d\x  \\
+M_p \dfrac{{\bf V}^{n+\frac24} - {\bf V}^{n+\frac{1}{4}}}{\triangle t} \cdot\Y 
+ {\bf I}_p \dfrac{  \bomega^{n+\frac24}- \bomega^{n+\frac{1}{4}}}{\triangle t} \cdot \bxi \\
 = (1-\dfrac{\rho_f}{\rho_s}) M_p {\bf g}\cdot\Y   +< \blambda^{n+\frac24},\
\bv-\Y-\bxi \times {\overrightarrow{\G^{n+\frac14}\x}}>_{\Lambda^{n+\frac14}_{h}}, \\ 
\forall \bv \in \W_{0h}, \ \Y \in\R^3, \ \bxi \in\R^3,
\end{cases} \label{eqn:34}\\
&&\hskip -12pt \into q\bnabla\cdot\bu^{n+\frac24} d\x=0, \ \forall q\in
L^2_{h},\\
&&\hskip -12pt < \bmu, \bu^{n+\frac24}-{\bf V}^{n+\frac24}-\bomega^{n+\frac24}\times {\overrightarrow{\G^{n+\frac14} \x}} >_{\Lambda^{n+\frac14}_{h}} = 0, \ \forall
\bmu \in \Lambda^{n+\frac14}_{h}.
\label{eqn:35}
\end{eqnarray}

\item We then compute  $\A^{n+\frac{3}{4}}$ via the solution of
\begin{equation}
\begin{cases}
\displaystyle \int_{\Omega}  \dfrac{\partial\A(t)}{\partial t} : \bs\, d \x + \int_{\Omega} (\bu^{n+\frac24} \cdot \bnabla)
 \A(t):\bs \, d\x =0, \forall \bs \in \V_{\A_{0h}},\\
\displaystyle \A(t^n)=\A^n, \,\, where\ \, \A^n{(\A^n)}^t=\bC^{n},\\
\displaystyle \A(t) \in \V^{n+1}_{\A_{Lh}},\, t \in [t^n,t^{n+1}],
\end{cases} \label{eqn:27}
\end{equation}
and set  $\A^{n+\frac{3}{4}}=\A(t^{n+1})$.

\item Finally we obtain $\A^{n+1}$ via the solution of
\begin{equation}
\begin{cases}
\displaystyle \int_{\Omega}(
 \dfrac{\A^{n+1}-\A^{n+\frac{3}{4}}}{\triangle t} 
-(\bnabla \bu^{n+\frac24})\A^{n+1}
+\dfrac{1}{2\lambda_1}\A^{n+1}): \bs \, d\x =0,\\
\displaystyle \forall \bs \in \V_{\A_{0h}};
\A^{n+1} \in \V^{n+1}_{\A_{Lh}},
\end{cases} \label{eqn:28}
\end{equation}
and set 
\begin{equation}
\bC^{n+1}=\A^{n+1}(\A^{n+1})^t +\dfrac{ \triangle t } {\lambda_1}\I, \ \text{and then} \ \bC^{n+1}=\I \ \text{in} \ B^{n+\frac14}. \label{eqn:29}
\end{equation}
Set ${\bf G}^{n+1}={\bf G}^{n+\frac14}$, ${\bf V}^{n+1}={\bf V}^{n+\frac24}$, and $\bomega^{n+1}=\bomega^{n+\frac24}$.

\end{enumerate}
In (\ref{eqn:27})-(\ref{eqn:29}),  the space $\V^{n+1}_{\A_{Lh}}$ is  $\V_{\A_{L_h}(t^{n+1})}$,  $\V_{\A_{L_h}(t)}$ and $\V_{\A_{0h}}$
being defined {similarly} to  $\V_{\bC_{L_h}(t)}$ and $\V_{\bC_{0h}}$. 
The multiplier space  $\Lambda_h^{n+\frac14}$ in (\ref{eqn:34})-(\ref{eqn:35}) is defined according to the position of $B^{n+\frac14}$. 

\begin{remark}
When simulating the motion of balls in a Newtonian fluid, we skip  problems (\ref{eqn:27}), (\ref{eqn:28}) and (\ref{eqn:29}) in 
algorithm (\ref{eqn:30})-(\ref{eqn:29}) and set the  elastic viscosity of the fluid to zero.
\end{remark}

\subsection{On the solution of the subproblems}

At the steps 3 and 4 {of}  algorithm (\ref{eqn:30})-(\ref{eqn:29}), we have considered the equations  {verified by} $\A$ 
instead {of those verified by the} conformation tensor $\bC$ due to the use of a factorization approach (e.g., see \cite{Hao2009} for  details).  
In the implementation, this kind of the Lozinski-Owens' scheme relies on the matrix factorization $\bC = \A\A^T$ of the conformation tensor, and then on a 
reformulation in terms of $\A$ of the time dependent equation modelling the evolution of $\bC$, providing 
automatically that $\bC$ is at least positive semi-definite (and symmetric). The matrix factorization 
based method introduced in \cite{lozinski2003} has been applied, via an operator splitting 
scheme coupled to a FD/DLM method, to the simulation of two-dimensional particulate flows 
of Oldroyd-B  and FENE fluids in \cite{Hao2009, Pan2017}.  

The equation (\ref{eqn:27}) is a pure advection problem.  We solve this equation by a wave-like equation method  (see, e.g., \cite{RG2003}, \cite{Dean1997}, 
and \cite{Hao2009} (p. 102))  which is a numerical dissipation free  explicit method. 
Since the advection problem is decoupled from the other ones, we can choose a proper sub-time step so that the CFL condition is satisfied.  
Problem (\ref{eqn:27})  gives a simple equation at each grid point which can be solved easily if we use trapezoidal quadrature rule to compute the integrals.
The value of $\bnabla \bu^{n}$ at each interior grid node is obtained by the averaged value of those values computed in all tetrahedral elements 
having the grid node as a vertex, {however} for  the grid node on $\Gamma$ it is obtained by applying linear extrapolation via the
values  of two neighboring interior nodes as discussed in \cite{Whiteman1991}.

Problem  (\ref{eqn:30})-(\ref{eqn:33}) is just a system of ordinary differential equations. 
They are solved using the forward Euler method with a sub-time step   to predict 
the translation velocity of the mass center and then  the position of the mass center.
But for the two ball interaction in a bounded shear flow,  we have applied the following approach developed in \cite{Guo2017} to predict the
ball positions: For the interaction during the two ball encounter, 
we have to impose a minimal gap of size $ch$ between the balls where $c$ is some constant between 0 and 1, $h$ being the 
mesh size of the velocity field. Then, when advancing the two ball mass centers, we proceed as follows at each sub-cycling time step: 
(i) we do nothing if the gap between the two balls at the new position is greater or equal than $ch$,
(ii) if the gap size of the two balls at the new position is less than $ch$, we do not advance the balls directly; but instead we first move the ball centers
in the direction perpendicularly to the line joining the previous centers, and then move them in the direction parallel to the line joining the previous centers, 
and make sure  that the gap size is no less than $ch$. For all the simulations reported in this article, relying on this strategy, we took $h/16$ as minimal gap size.

In system (\ref{eqn:34})--(\ref{eqn:35}), there are two multipliers: namely $p$ and  $\blambda$. 
We have solved this system via an Uzawa-conjugate gradient method driven by both 
multipliers (an one shot method, similar to those discussed in, e.g., \cite{Pan2015, RG1994,RG1995}). The general problem is as follows:

Find $\bu\in \W_h$, $\bu=\g_{0}$  on $\Gamma$, $p\in L^2_{0h}$, $\blambda\in \Lambda_{h}$, 
${\bf V}\in\R^3$, $\bomega\in\R^3$ so that 
\begin{eqnarray}
&&\begin{cases}
-\into p \bnabla \cdot\bv d\x   + \mu \into \bnabla\bu:\bnabla \bv\, d\x   
+M_p \dfrac{{\bf V} -{\bf V}_0}{\triangle t} \cdot\Y 
+\dfrac{ {\bf I}_p \bomega- \bomega_0}{\triangle t} \cdot \bxi \\
\hskip 10pt  = (1-\dfrac{\rho_f}{\rho_s}) M_p {\bf g}\cdot\Y  + < \blambda,\
\bv-\Y-\bxi \times {\Gx}>_{\Lambda_h} + \into {\bf F}\cdot \bv\, d\x, \\ 
\forall \bv \in \W_{0h}, \ \Y \in\R^3, \ \bxi \in\R^3,
\end{cases} \label{eqn:36} \\
&& \into q\bnabla\cdot\bu d\x=0, \ \forall q\in
L^2_{h},\label{eqn:37}  \\
&&  < \bmu, \bu-{\bf V}-\omega\ {\Gx} >_{\Lambda_h} = 0, \ \forall
\bmu \in \Lambda_{h}.
\label{eqn:38}
\end{eqnarray}
To solve system (\ref{eqn:36})--(\ref{eqn:38}) we employed the following Uzawa-conjugate gradient algorithm operating in the space 
$L^2_{0h}\times\Lambda_{h}$:

\vskip 2ex

{\it $p^0 \in L^2_{0h}$ and  $\blambda^0 \in\Lambda_{h} $ are given; }
\vskip 1ex
\noindent{\it solve}
\begin{eqnarray}
&&\begin{cases}
\mu \into \bnabla\bu^0:\bnabla \bv\, d\x = 
\displaystyle\int_{\Omega} p^0\ \nabla \cdot \bv \, d\x+< \blambda^0,\ \bv>_{\Lambda_h}+ \into {\bf F}\cdot \bv\, d\x, \\ 
\forall \bv \in \W_{0h}; \ \bu^0 \in \W_{h}, \ \bu=\g_{0h} \ on \ \Gamma, 
\end{cases} \label{eqn:s39}\\
&&M_p \dfrac{{\bf V}^0 -{\bf V}_0}{\triangle t} \cdot\Y=(1-\dfrac{\rho_f}{\rho_s}) M_p {\bf g}\cdot\Y
- < \blambda^0,\ \Y>_{\Lambda_h},\ \forall \Y \in\R^3, \label{eqn:s40}\\
&&\dfrac{{\bf I}_p\bomega^0-\bomega_0}{\triangle t}\cdot \bxi = - < \blambda^0,\bxi \times  {\Gx}>_{\Lambda_h}, \ \forall \bxi \in\R^3, \label{eqn:s41}
\end{eqnarray}
{\it and then compute}
\begin{equation}
{\rm g}_1^0 = \nabla \cdot \bu^0; \label{eqn:s42}
\end{equation}
{\it next solve }
\begin{equation}
\begin{cases}
\g_2^0 \in \Lambda_h,\\
<\bmu, \g_2^0>_{\Lambda_{h}}= <\bmu, \ \bu^0 - {\bf V}^0-\bomega^0\times {\Gx}>_{\Lambda_h}, 
\ \forall \bmu \in \Lambda_{h},
\end{cases}\label{eqn:s43}
\end{equation}
{\it and set}
\begin{equation}
{\rm w}_1^0={\rm g}_1^0, \ \w_2^0=\g_2^0.\label{eqn:s44}
\end{equation}

{\it Then for $k \ge 0$, assuming that $p^k$, $\blambda^k$, $\bu^k$, $\V^k$, $\bomega^k$,  
${\rm g}_1^k$, $\g_2^k$, ${\rm w}_1^k$ and $\w_2^k$ are known, compute $p^{k+1}$,  $\blambda^{k+1}$,
$\bu^{k+1}$, $\V^{k+1}$, $\bomega^{k+1}$, ${\rm g}_1^{k+1}$, $\g_2^{k+1}$, ${\rm w}_1^{k+1}$
$\w_2^{k+1}$  as follows:}
\vskip 1ex 
\noindent {\it solve:}
\begin{eqnarray}
&&\begin{cases}
\mu \into \bnabla{\overline \bu}^{\, k}:\bnabla \bv\, d\x = 
\displaystyle\int_{\Omega} {\rm w}_1^k\ \nabla \cdot \bv \, d\x +< \w_2^k,\ \bv>_{\Lambda_h}, \\ 
\forall \bv \in \W_{0h}; \ {\overline \bu}^{\, k} \in \W_{0h},
\end{cases} \label{eqn:s45}\\
&&M_p \dfrac{{\overline {\bf V}}^{\, k}}{\triangle t} \cdot\Y=- < \w_2^k,\ \Y>_{\Lambda_h},\ \forall \Y \in\R^3,\label{eqn:s46}\\
&&{\bf I}_p\dfrac{{\overline \bomega}^{\, k}}{\triangle t} \cdot \bxi = - < \w_2^k,\bxi \times {\Gx}>_{\Lambda_h},  \ \forall \bxi \in\R^3,\label{eqn:s47}
\end{eqnarray}
{\it and then compute}
\begin{equation}
{\overline {\rm g}}_1^{\, k} = \nabla \cdot {\overline \bu}^{\, k}; \label{eqn:s48}
\end{equation}
{\it next solve}
\begin{equation}
\begin{cases}
{\overline \g}_2^{\, k} \in \Lambda_h,\\
<\bmu, {\overline \g}_2^{\, k}>_{\Lambda_{h}} = <\bmu, \ {\overline \bu}^{\, k} - {\overline {\bf V}}^{\, k}
-{\overline \bomega}^{\, k}\times {\Gx}>_{\Lambda_h}, \ \forall \bmu \in \Lambda_{h},
\end{cases}\label{eqn:s49}
\end{equation}

\noindent{\it and compute} 
\begin{equation}
\displaystyle 
\rho_k=\dfrac{\int_{\Omega} |{\rm g}_1^k|^2 \, d\x + <\g_2^k, \g_2^k>_{\Lambda_{h}}}
{\int_{\Omega} {\overline {\rm g}_1}^{k}  {\rm w}_1^k \, d\x+<{\overline \g}_2^{\, k}, \w_2^k>_{\Lambda_{h}}},
\label{eqn:s50}
\end{equation}
{\it and}
\begin{eqnarray}
&&p^{k+1}=p^k -\rho_k {\rm w}_1^k, \label{eqn:s51}\\
&&\blambda^{k+1}=\blambda^k - \rho_k \w_2^k, \label{eqn:s52} \\
&&\bu^{k+1}=\bu^k -\rho_k {\overline \bu}^{\, k}, \label{eqn:53}\\
&&\V^{k+1}=\V^k - \rho_k  {\overline \V}^{\, k}, \label{eqn:s54}\\
&&\bomega^{k+1}=\bomega^k - \rho_k  {\overline \bomega}^{\, k},\label{eqn:s55}\\
&&{\rm g}_1^{k+1}={\rm g}_1^k - \rho_k  {\overline {\rm g}}_1^{\, k}, \label{eqn:s56}\\
&&\g_2^{k+1}=\g_2^k - \rho_k  {\overline \g}_2^{\, k}.\label{eqn:s57}
\end{eqnarray}
\vskip 2ex

\noindent{\it If} 
\begin{equation}
\dfrac{\int_{\Omega} |{\rm g}_1^{k+1}|^2 \, d\x + <\g_2^{k+1}, \g_2^{k+1}>_{\Lambda_{h}}}
{\int_{\Omega} |{\rm g}_1^0|^2 \, d\x + <\g_2^0, \g_2^0>_{\Lambda_{h}}} \le tol
\label{eqn:s58}
\end{equation}
{\it take $p=p^{k+1}$, $\blambda=\blambda^{k+1}$, $\bu=\bu^{k+1}$, $\V=\V^{k+1}$,
$\bomega=\bomega^{k+1}$; else, compute}
\begin{equation}
\displaystyle 
\gamma_k=\dfrac{\int_{\Omega} |{\rm g}_1^{k+1}|^2 \, d\x + <\g_2^{k+1}, \g_2^{k+1}>_{\Lambda_{h}}}
{\int_{\Omega} |{\rm g}_1^k|^2 \, d\x + <\g_2^k, \g_2^k>_{\Lambda_{h}}}
\label{eqn:s59}
\end{equation}
{\it and set} 
\begin{eqnarray}
&&{\rm w}_1^{k+1}={\rm g}_1^{k+1}+\gamma_k {\rm w}_1^k,\label{eqn:s60}\\
&&\w_2^{k+1}=\g_2^{k+1}+\gamma_k \w_2^k.\label{eqn:s61}
\end{eqnarray}
{\it Do $k \gets k+1$ and go back to} (\ref{eqn:s45}).

In this article, we took $tol=10^{-14}$.

\section{Numerical results}
\begin{figure}[thp!]
\begin{center}
\leavevmode
\epsfxsize=3.25in
\epsffile{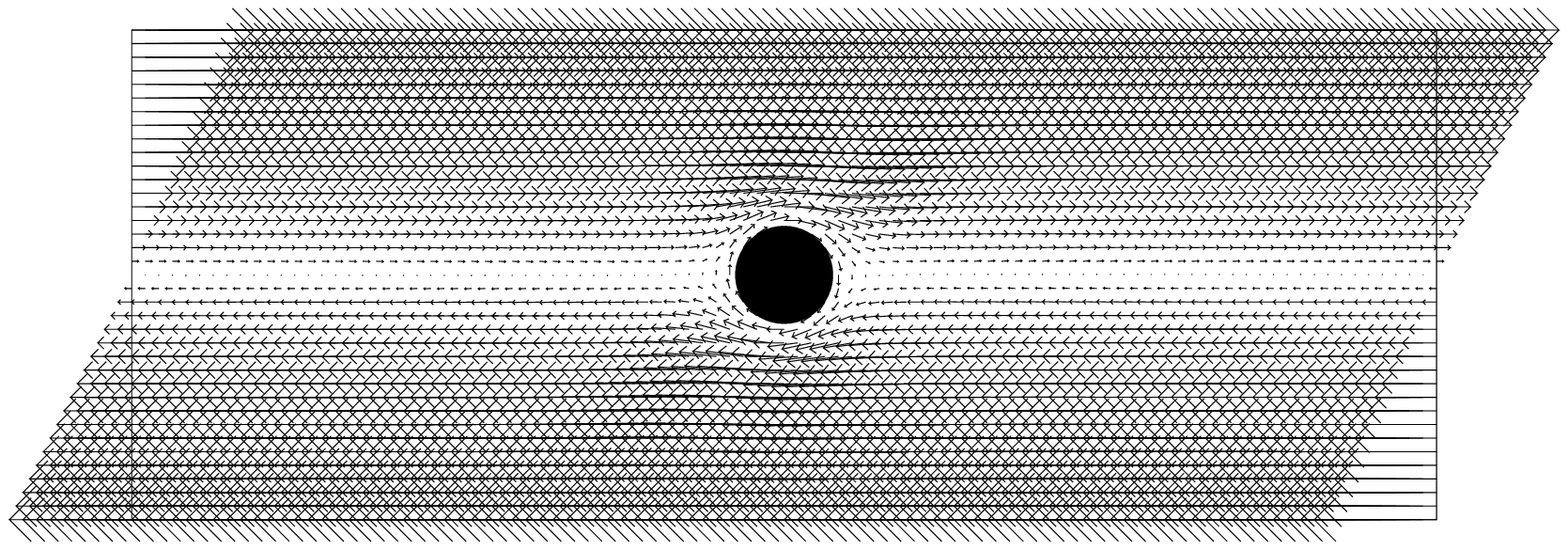}\\
\vskip 0.1in
\epsfxsize=3.25in
\epsffile{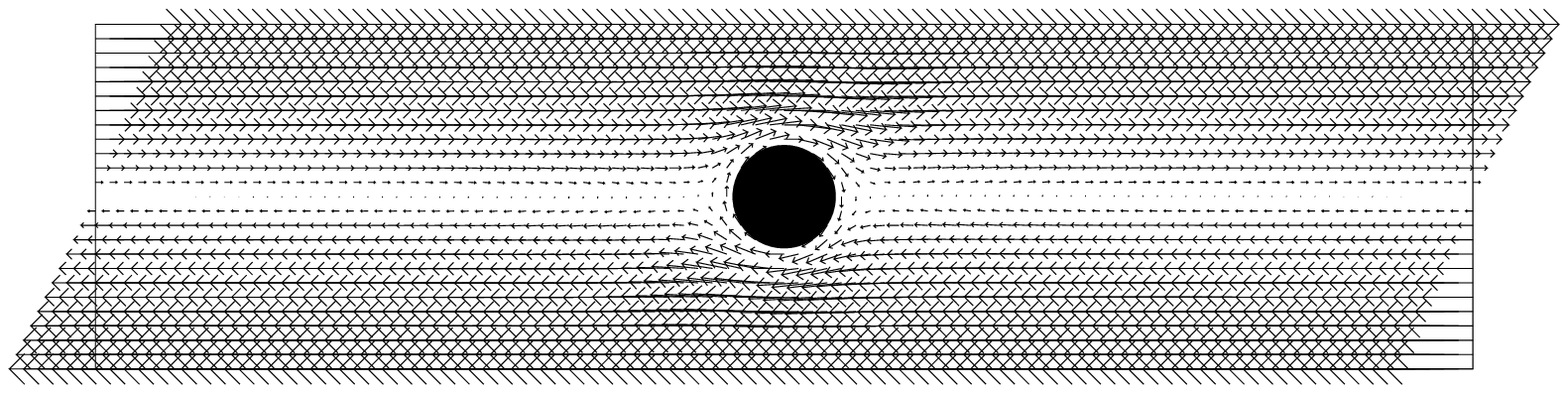}
\end{center}
\caption{Snapshots of the velocity field projected on the $x_1x_3$-plane  for  $K$ = 0.2 (top) and 0.3 (bottom). }\label{fig:3a}
\begin{center}
\leavevmode
\epsfxsize=3.25in
\epsffile{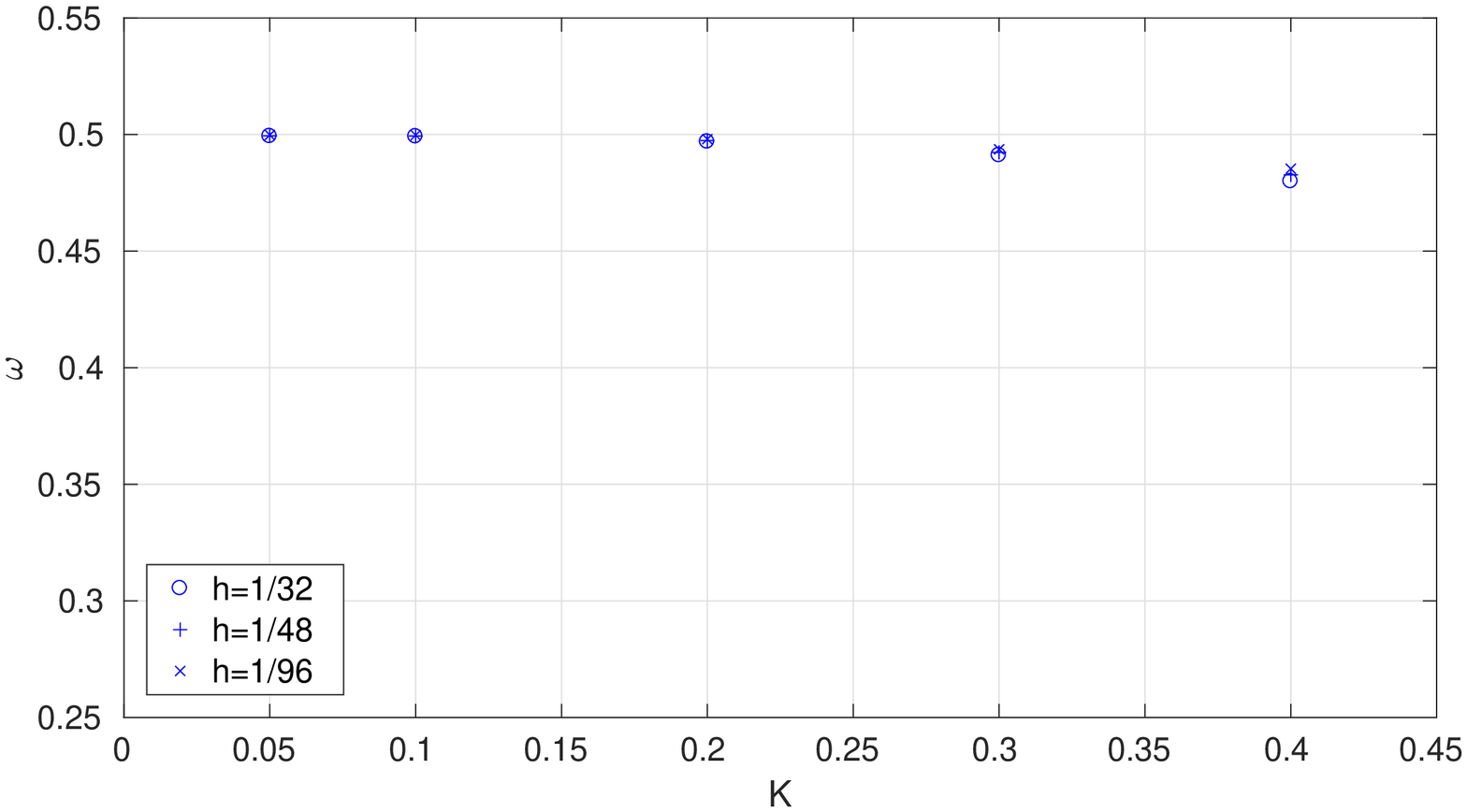}    \\
\epsfxsize=3.25in
\epsffile{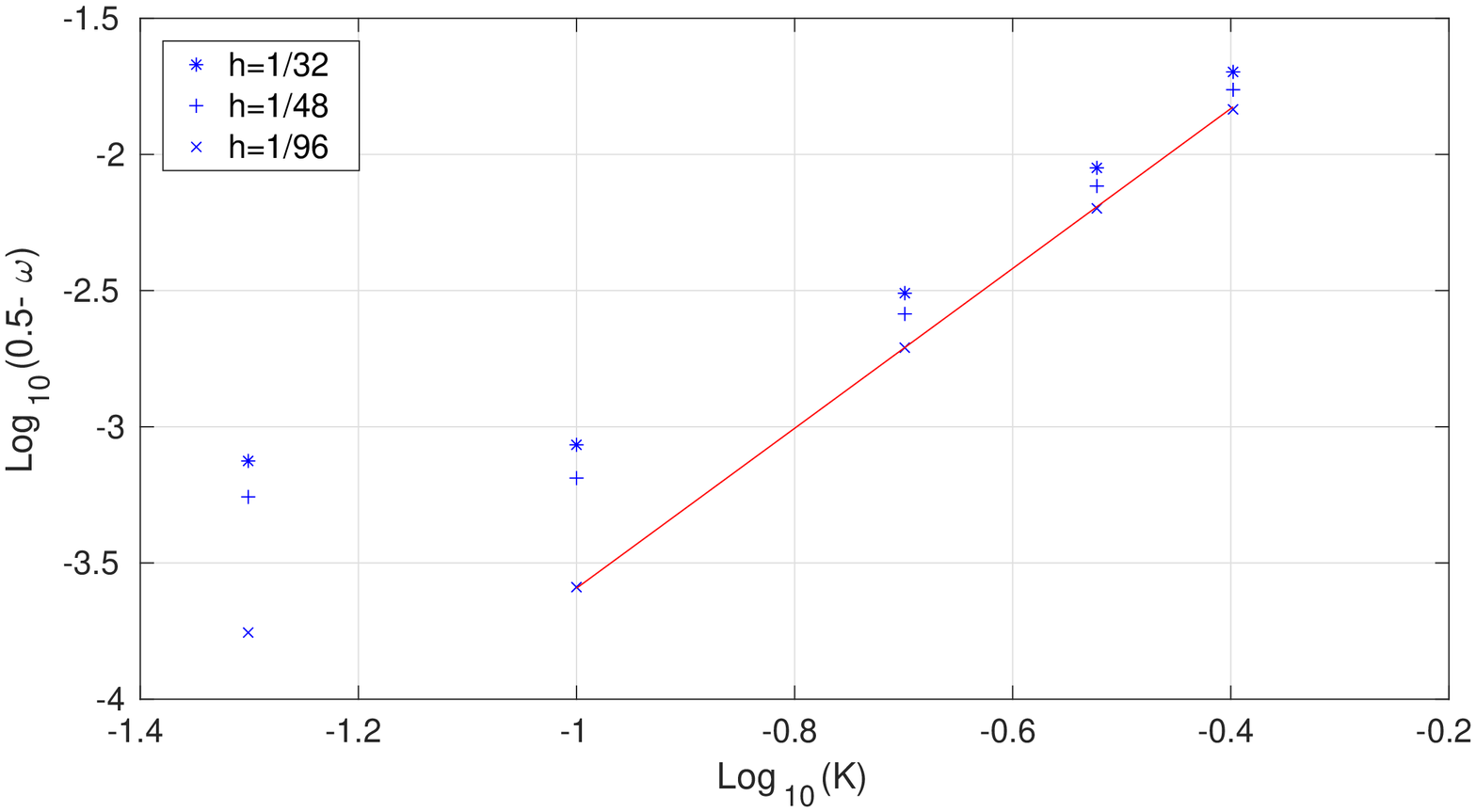}
\end{center}
\caption{ The rotating speed versus the confined ratio (top) and the log--log plot of the difference of the rotation velocity versus the confined ratio (bottom).
The solid line in the log--log plot {shows  the following power law effect} of the confined ratio:  $\omega=0.5-{0.22} K^{2.935}$ for $0.1 \le K \le 0.4$.}\label{fig:3b}
\end{figure}

\begin{figure} [tp!]
\begin{center}
\leavevmode
\epsfxsize=3.5in
\epsffile{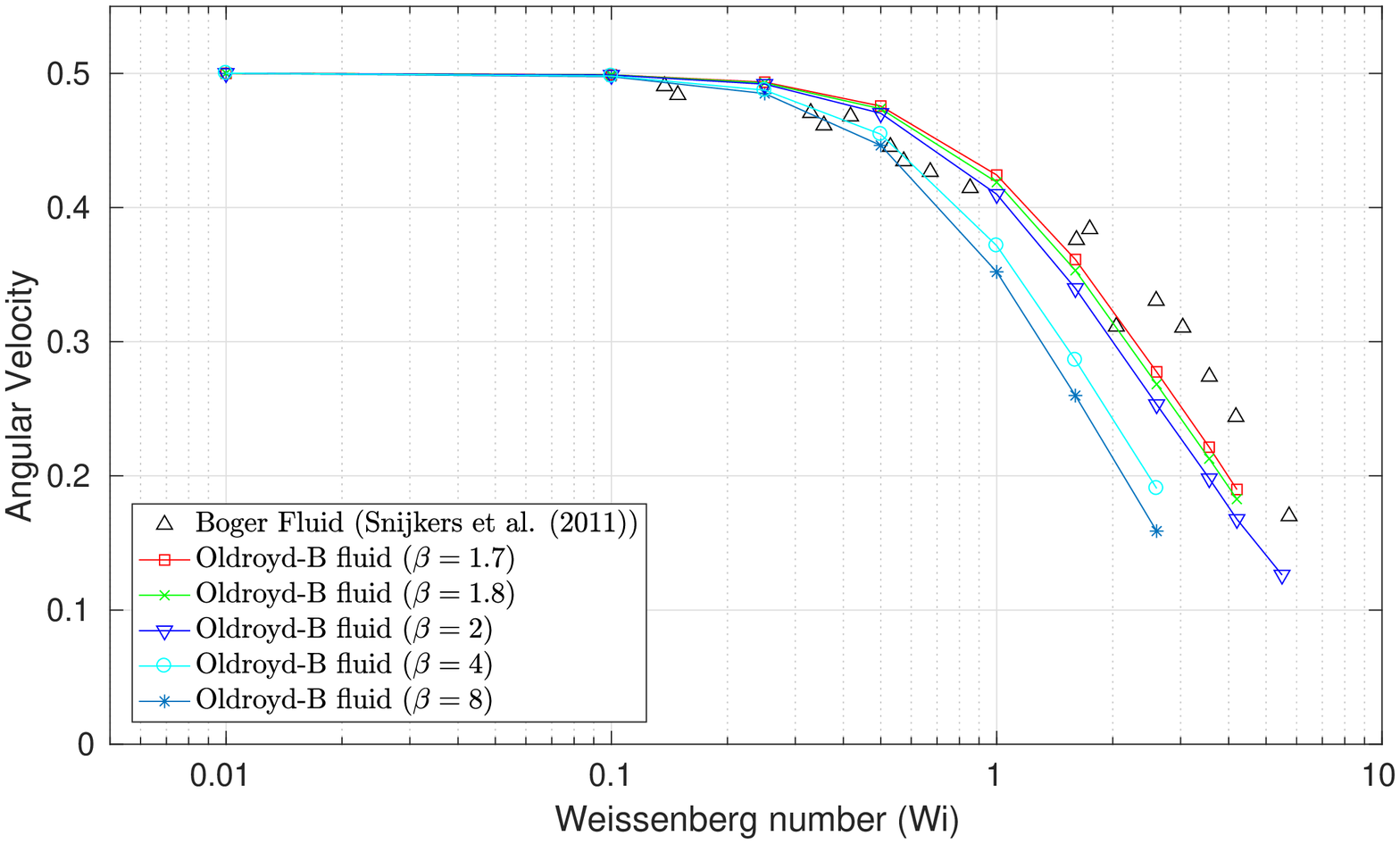} 
\end{center}
\caption{Comparison of the angular velocity of a single ball freely rotating  in an Oldroyd-B fluid with its 
mass center fixed at (0,0,0) for different values of {the} Weissenberg number Wi ($=\lambda_1\dot{\gamma}$).}\label{fig:4a}
\end{figure}

\subsection{A ball rotating in a bounded  shear flow}
 
We have considered first the {case} of a neutrally buoyant ball which is suspended and freely moving in a Newtonian fluid.
Its mass center is located at $(0,0,0)$ initially. The computational domain is $\Omega= (-2, 2) \times (-2, 2) \times (-H/2, H/2)$  
(i.e., $L_1 = L_2=4$), the values of $H$ being 0.75, 1, 1.5, 3, and 6. 
The ball radius $a$ is 0.15, while the fluid and particle densities are $\rho_f = \rho_s = 1$, the fluid viscosity being $\mu_f = 1$. 
The confined ratio is defined as $K = 2a/ H$ where $H$ is the distance between {the} two horizontal walls. 
The shear rate is fixed at $\dot{\gamma}= 1$ so that the velocity of the top wall (resp., bottom wall) is $U= H/2$
(resp., $-U = -H/2$).   
The mesh size for the velocity field is either  $h = 1/32$, 1/48,  or 96, the mesh size for the pressure is $2h$,
and the time step is  $\triangle t= 0.001$. Under creeping flow conditions, the rotating velocity of the ball with respect to the $x_2$-axis 
is $\dot{\gamma}/2=0.5$  in an unbounded shear flow according to the associated Jeffery's solution \cite{Jeffery1922}. 
Snapshots of the velocity field projected on the $x_1x_3$-plane for the cases $K=$ 0.2 and 0.3, computed with  $h=1/96$, are shown in Fig. \ref{fig:3a},
The plot of the rotation speed versus the confined ratio being presented in Fig. \ref{fig:3b}. The computed angular speeds  for $K=0.05$ and 0.1 
are in a good agreement with Jeffery's solution. The confined ratio {affects} the rotation speed {as visualized in Fig. \ref{fig:3b} 
where the solid line in the log--log plot shows for $\omega$ a confined ratio power law dependence given (approximately) by}
$\omega=0.5- 0.22 K^{2.935}$ for $0.1 \le K \le 0.4$.  For all the numerical simulations considered in this article, we assume that all 
dimensional quantities are in the {physical} CGS units.

For the cases of a single ball freely rotating in an Oldroyd-B fluid  with its mass center fixed at (0,0,0),  we have  considered 
different values of the relaxation time $\lambda_1$.  The ball radius $a$ is 0.1, while the fluid and particle densities are $\rho_f = \rho_s = 1$, 
the fluid viscosity being $\mu_f = 1$.  The computational domain is $\Omega=(-1.5,1.5) \times (-1.5,1.5) \times (-1.5,1.5)$. Then the 
blockage ratio is $K=1/15$ (same as the one used in  \cite{Snijkers2011}). The mesh size for the velocity field is  $h =1/64$, the mesh size for 
the pressure is $2h$, and the time step is  $\triangle t= 0.001$.  The rotating velocities reported in Fig. \ref{fig:4a} are in a good agreement 
with those reported in \cite{Snijkers2011}.   We have also considered different {values of} the retardation times, {namely} $\lambda_2=\lambda_1/\beta$ 
{with} $\beta=1.7$, 1.8,  2, 4, and 8.  Our numerical results {shown in Fig. \ref{fig:4a}} suggest that the retardation {time affects also the} 
rotating speed. The reasonable range of the value of $\beta=\frac{\lambda_1}{\lambda_2}$ is 
about between 1.7 and 2 when comparing with the experimental results of the rotating velocity in a Boger fluid reported in \cite{Snijkers2011}.
 
\subsection{Two balls interacting in a two wall driven bounded shear flow}

\begin{figure}[t!]
\begin{center}
\leavevmode
\epsfysize=1.1in
\epsffile{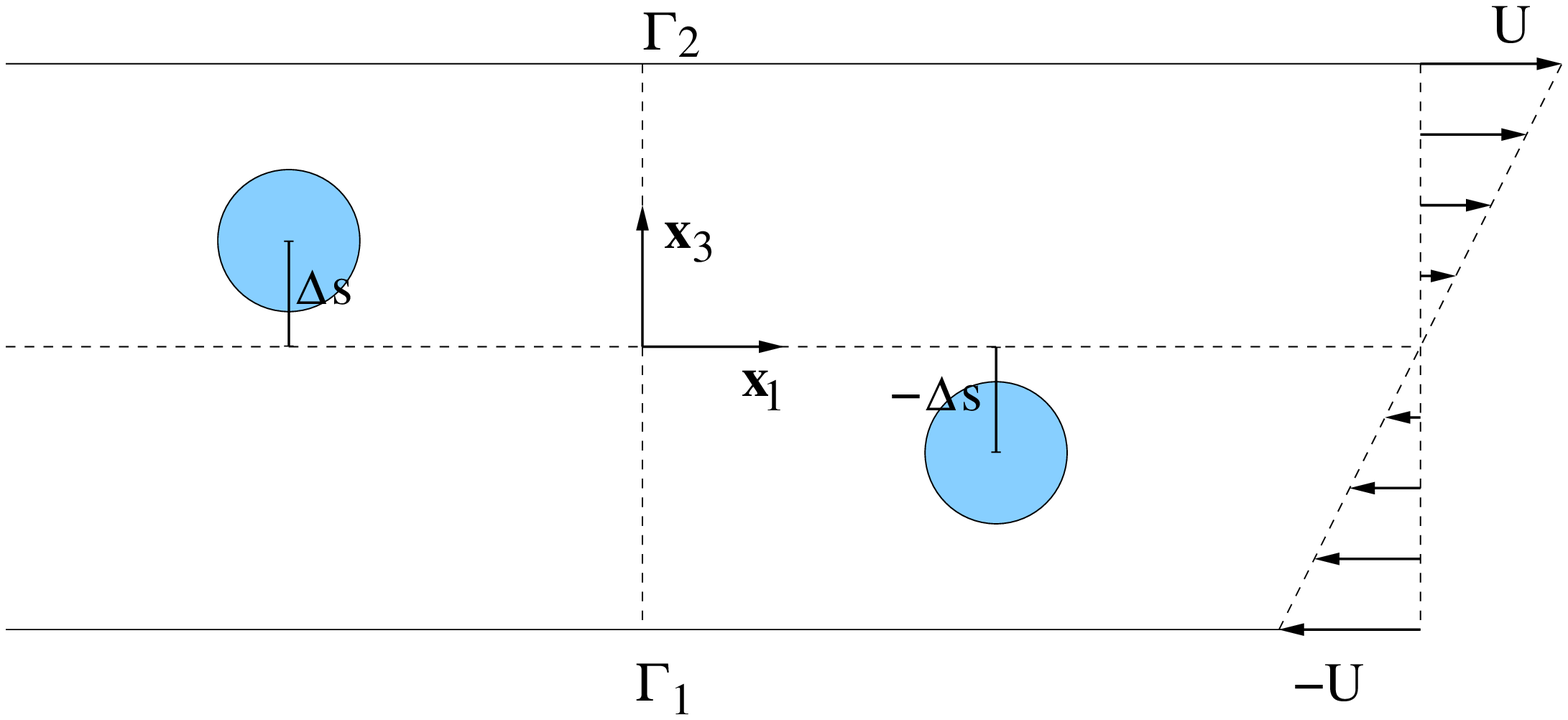}\hskip10pt 
\epsfysize=1.1in
\epsffile{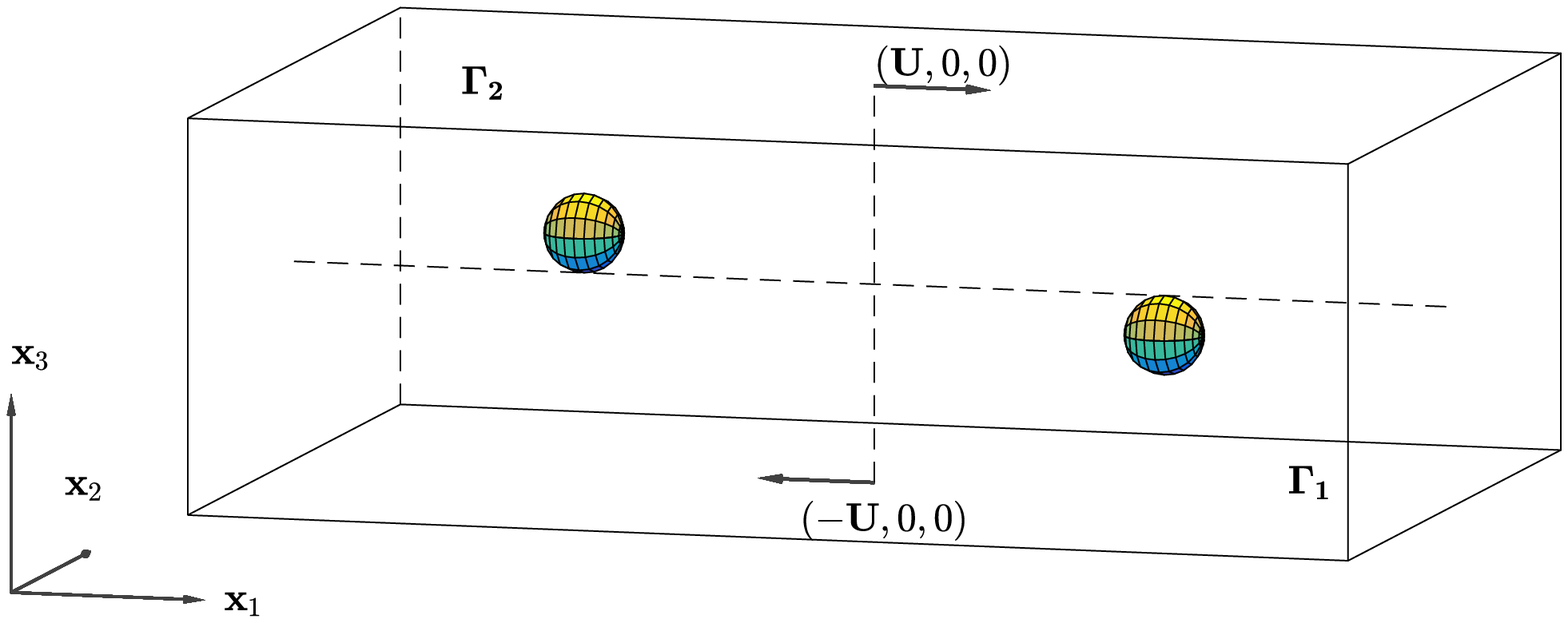}
\end{center}
\caption{Two balls interacting  in a bounded shear flow.}\label{fig:6a}
\end{figure}

\begin{figure}[tp!]
\begin{center}
\leavevmode
\epsfxsize=6.in
\epsffile{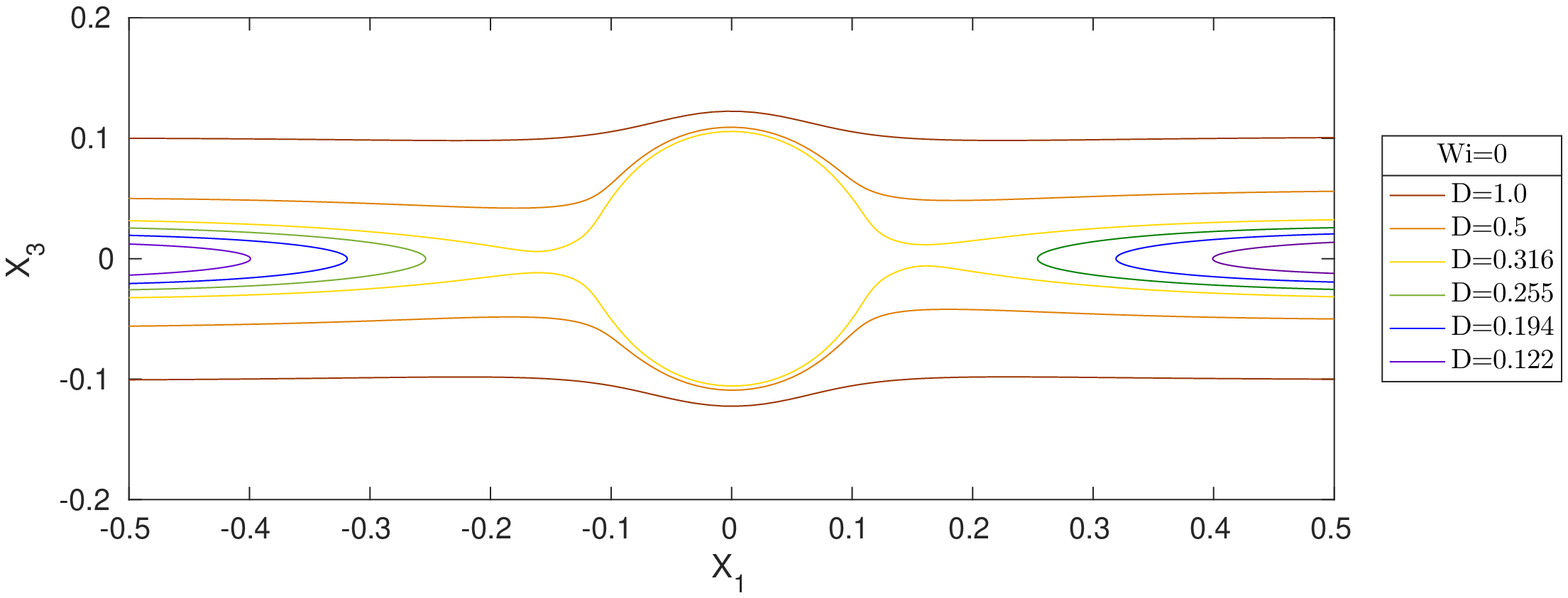}
\end{center}
\caption{Projected trajectories of the two ball mass centers in a  bounded shear flow for Wi=0 where the higher ball (initially located 
above $x_3=0$ and at $x_1=-0.5$) moves from the left to the right and the lower ball  (initially located  below $x_3=0$ and at $x_1=0.5$) moves 
from the right to the left: (a)  the balls pass over/under for $D=$1.0, 05, and 0.316, and (b) the  balls {swap} for $D=$0.255, 0.194, and 0.122. }\label{fig:7a}
\end{figure}

\begin{figure}[th!]
\begin{center}
\leavevmode
\epsfxsize=6.0in
\epsffile{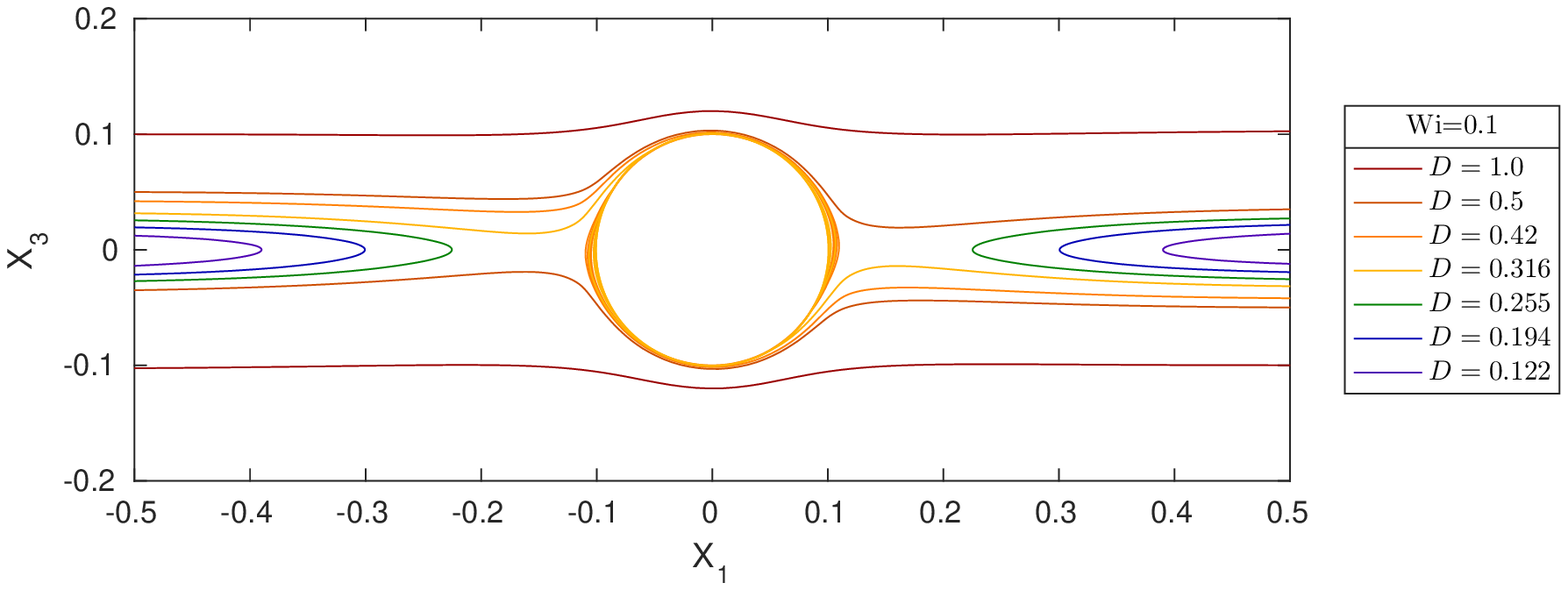}
\end{center}
\caption{Projected trajectories of the two ball mass centers in a  bounded shear flow for Wi=0.1 where the higher ball (initially located 
above $x_3=0$ and at $x_1=-0.5$) moves from the left to the right and the lower ball  (initially located  below $x_3=0$ and at $x_1=0.5$) moves 
from the right to the left: (a)  the  balls pass over/under for $D=$ 0.1 and 0.5,
(b) the  balls {swap} for $D=$ 0.255, 0.194 and 0.122, and (c) the  balls chain and then tumble for $D=$0.316 and 0.42. }\label{fig:8a}
\end{figure}

\begin{figure}[tp!]
\begin{center}
\leavevmode
\epsfxsize=6.0in
\epsffile{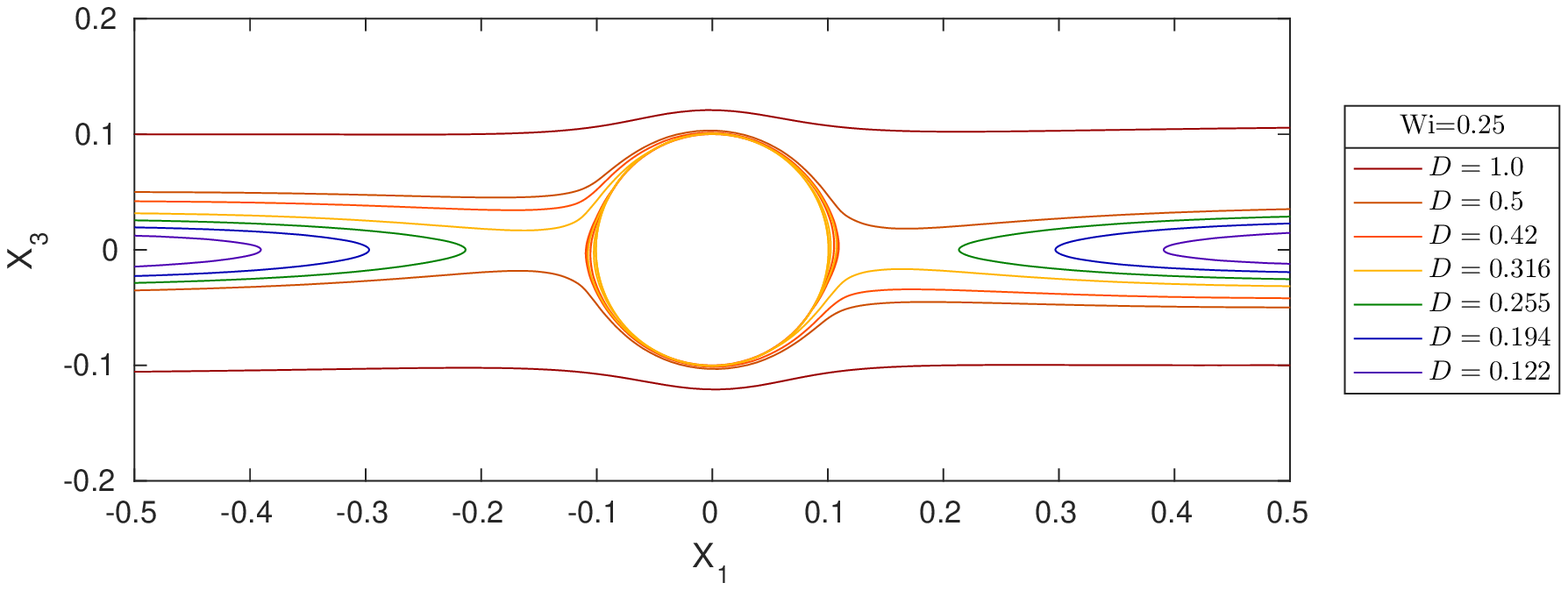}
\end{center}
\caption{Projected trajectories of the two ball mass centers in a  bounded shear flow for Wi=0.25 where the higher ball (initially located 
above $x_3=0$ and at $x_1=-0.5$) moves from the left to the right and the lower ball  (initially located  below $x_3=0$ and at $x_1=0.5$) moves 
from the right to the left: (a)  the balls pass over/under for $D=$ 0.1 and 0.5,
 (b) the balls {swap} for $D=$ 0.255, 0.194 and 0.122, and (c) the balls chain and  tumble for $D=$ 0.316 and 0.42.}\label{fig:9a}
\end{figure}

\begin{figure}[tp!]
\begin{center}
\leavevmode
\epsfxsize=6.in
\epsffile{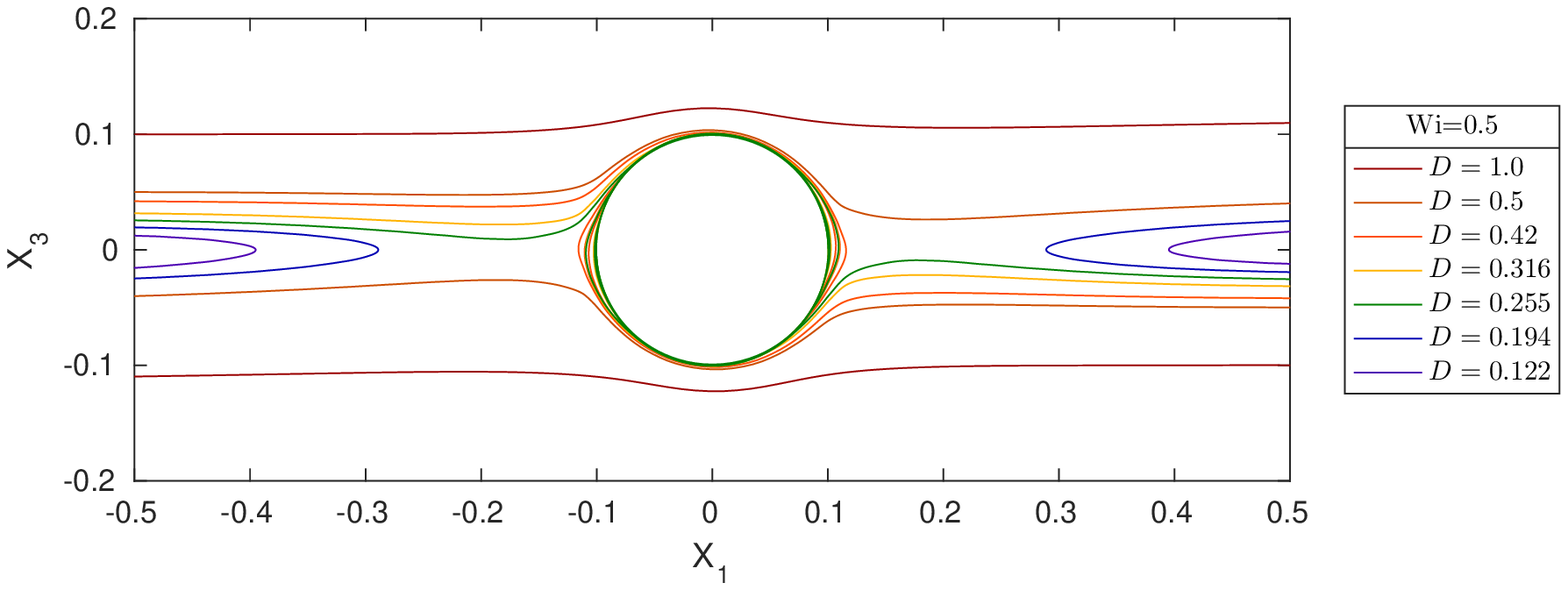}
\end{center}
\caption{Projected trajectories of the two ball mass centers in a bounded shear flow for Wi=0.5 where the higher ball (initially located 
above $x_3=0$ and at $x_1=-0.5$) moves from the left to the right and the lower ball  (initially located  below $x_3=0$ and at $x_1=0.5$) moves 
from the right to the left: (a)  the balls pass over/under  for $D=$ 0.1 and 0.5,
 (b) the balls {swap} for $D=$ 0.194 and 0.122, and (c) the balls chain and  tumble for $D=$ 0.255, 0.316
 and 0.42. }\label{fig:10a}
\end{figure}

\begin{figure}[tp!]
\begin{center}
\leavevmode
\epsfxsize=6in
\epsffile{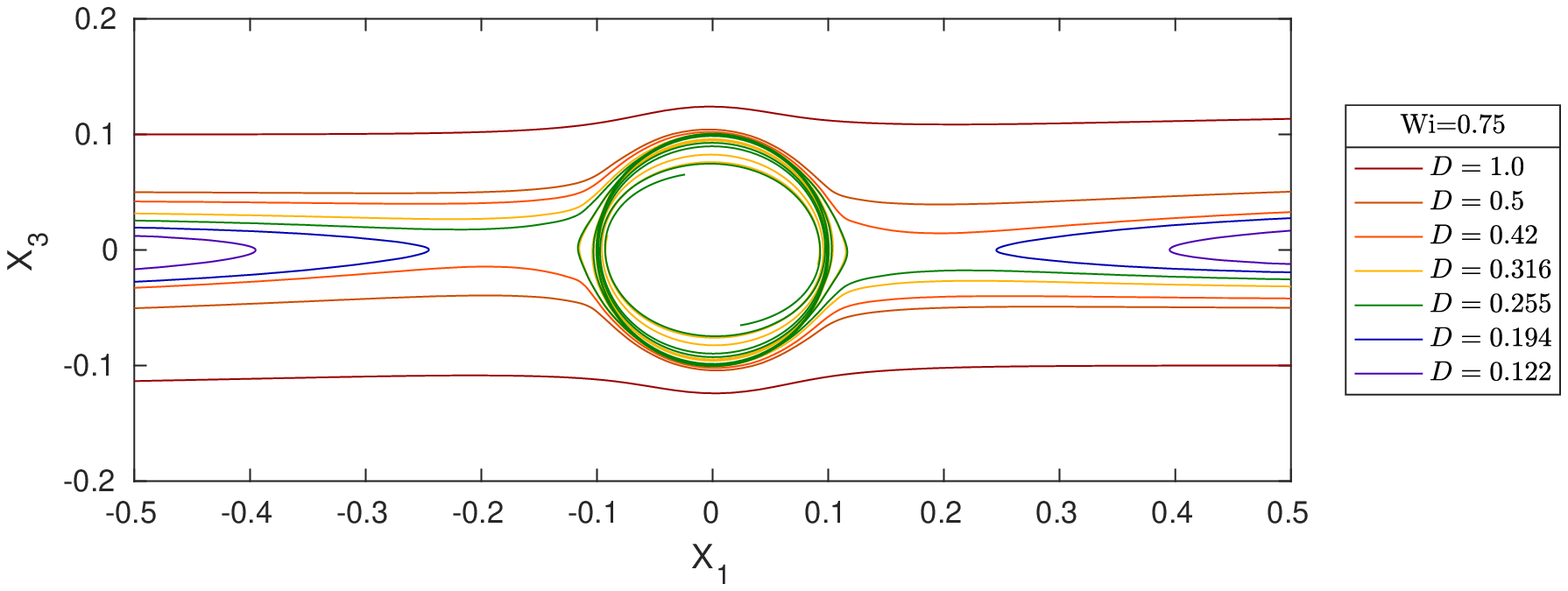}
\end{center}
\caption{Projected trajectories of the two ball mass centers in a bounded shear flow for Wi=0.75 where the higher ball (initially located 
above $x_3=0$ and at $x_1=-0.5$) moves from the left to the right and the lower ball  (initially located  below $x_3=0$ and at $x_1=0.5$) moves 
from the right to the left: (a)  the balls pass over/under  for $D=$ 0.1, 0.5 and 0.42,
(b) the balls {swap} for $D=$ 0.194 and 0.122, and (c) the balls tumble and then  kayak for $D=$ 0.255 and 0.316.}\label{fig:11a}
\end{figure}

\begin{figure}[tp!]
\begin{center}
\leavevmode
\epsfxsize=6in
\epsffile{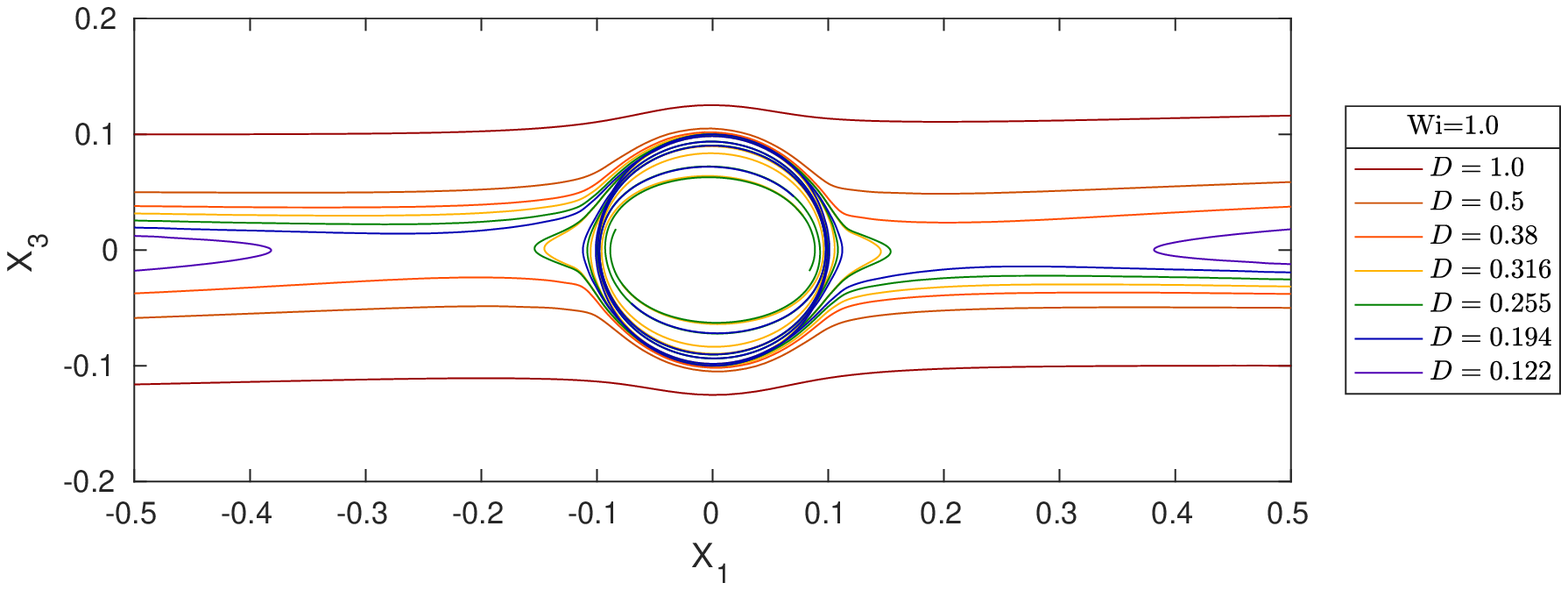}
\end{center}
\caption{Projected trajectories of the two ball mass centers in a  bounded shear flow for Wi=1 where the higher ball (initially located 
above $x_3=0$ and at $x_1=-0.5$) moves from the left to the right and the lower ball  (initially located  below $x_3=0$ and at $x_1=0.5$) moves 
from the right to the left: (a)  the balls pass over/under for $D=$ 0.1, 0.5 and 0.38
 (b) the balls {swap} for $D=$ 0.122, and (c) the balls tumble  and then  kayak for $D=$ 0.194, 0.255 and  0.316.}\label{fig:12a}
\end{figure}

In this section we  consider  the {case} of two balls of the same size interacting in a bounded shear flow as visualized in Fig. \ref{fig:6a}. 
The ball radii are $a=0.1$. The  fluid and ball densities are $\rho_f$= $\rho_s$ = 1, the viscosity being $\mu$ = 1. The relaxation time
$\lambda_1$ {takes the values} 0.1, 0.25, 0.5, 0.75 and 1, the retardation time being $\lambda_2=\lambda_1/8$. 
The computational domain  is $\Omega= (-1.5, 1.5) \times (-1,1) \times (-0.5, 0.5)$  (i.e., $L_1 =3$ and $L_2=2$). 
The shear rate is fixed at $\dot{\gamma} = 1$ so the velocity of 
the top wall is $U=0.5$,  the bottom wall {velocity}  being  $U=-0.5$.  The mass centers of the two balls are located on the shear plane at 
$(-d_0,0,\triangle s)$ and $(d_0,0,-\triangle s)$ initially, where $\triangle s$ varies and $d_0$ is  0.5.
The mesh size for the velocity field and the conformation tensor is $h = 1/48$, the mesh size for the pressure is $2h$,  the time step being
$\triangle t = 0.001$. The dimensionless initial vertical displacements from the ball center to the middle plane, namely $D=\triangle s/a$, 
are indicated in Figs  \ref{fig:7a} to  \ref{fig:12a}. The Weissenberg number is Wi=$\dot{\gamma}\lambda_1$.

When two balls move in  a bounded shear flow of a Newtonian fluid at Stokes regime,  the higher ball takes over the lower one and then both return to 
their initial heights for those  large  vertical displacements $D =$ 0.316, 0.5 and 1 as 
in Fig. \ref{fig:7a}.  
These two particle paths are called pass (or open) trajectories. But for smaller  vertical
displacements, $D =$ 0.122, 0.255 and 0.316, they first come close to each other 
and to the mid-plane between the two horizontal walls, then, the balls move away from 
each other and from the above mid-plane. These paths of the two particle are called return
trajectories.  Both kinds  are on the shear plane as shown in Fig.  \ref{fig:7a}
for Wi=0 (Newtonian case) and they are consistent with the results obtained in \cite{Zurita2007}.

\begin{figure}[tp!]
\begin{center}
\leavevmode
\epsfxsize=4in
\epsffile{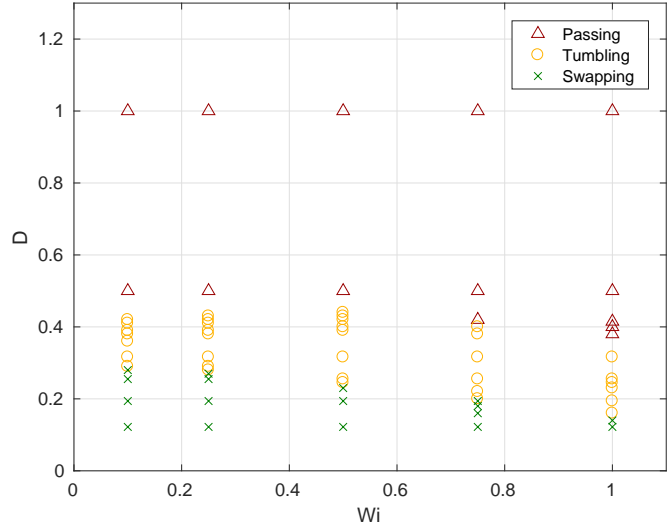}
\end{center}
\caption{Phase diagram for the motion of two balls based on the initial vertical displacement $D$ and {on the} Weissenberg number Wi in a 
bounded shear flow.}\label{fig:13a}
\end{figure}

\begin{figure}[tp!]
\begin{center}
\leavevmode
\epsfxsize=1.95in
\epsffile{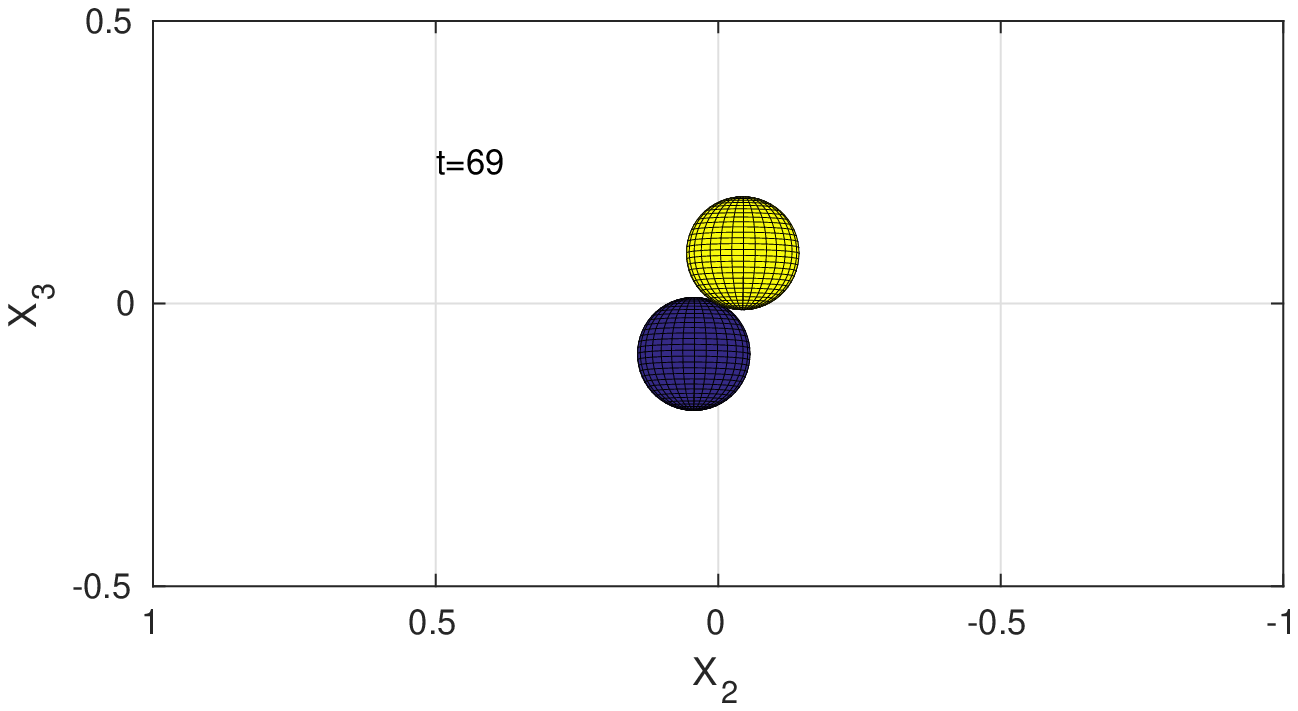}
\epsfxsize=1.95in
\epsffile{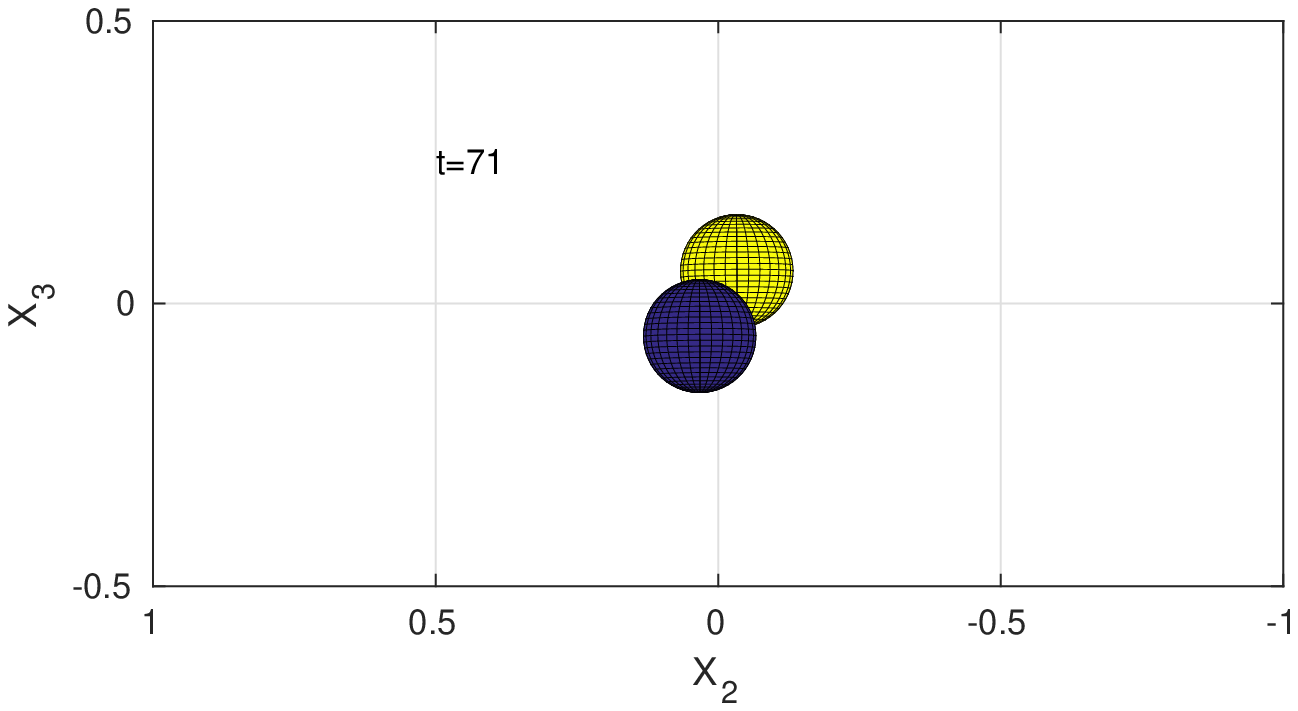}
\epsfxsize=1.95in
\epsffile{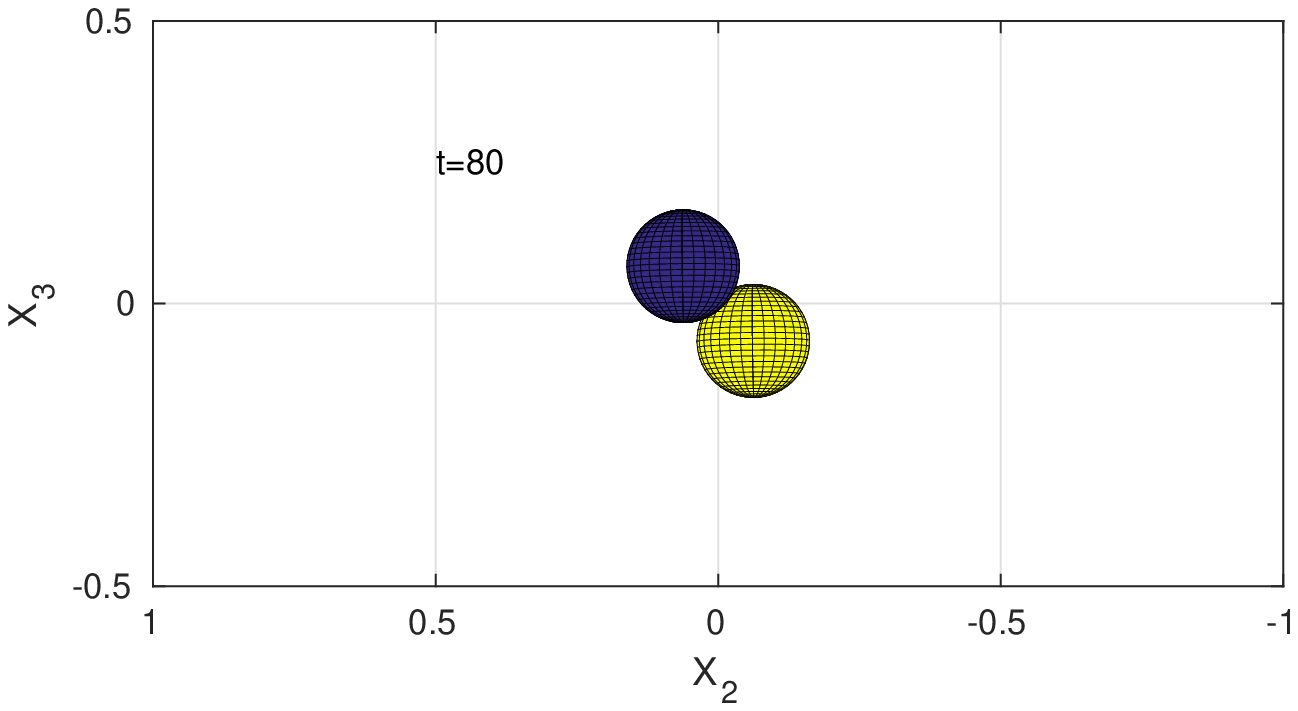}\\
\epsfxsize=1.95in
\epsffile{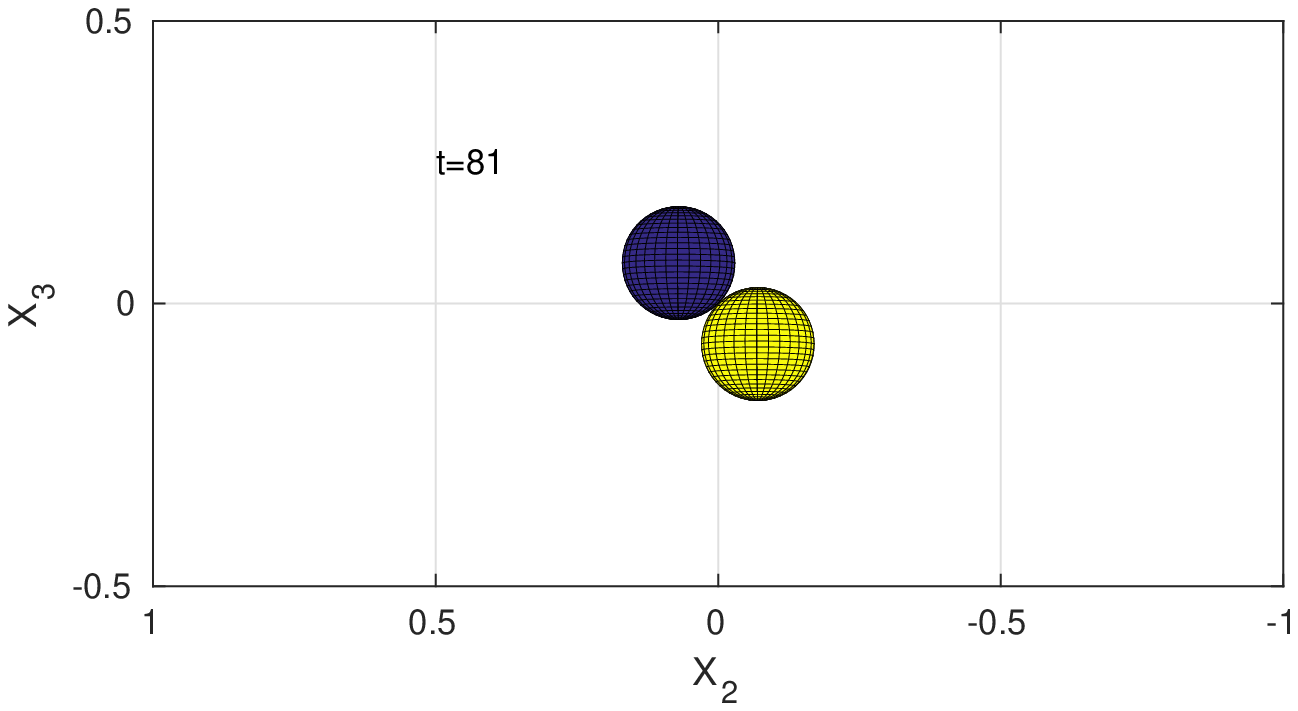}
\epsfxsize=1.95in
\epsffile{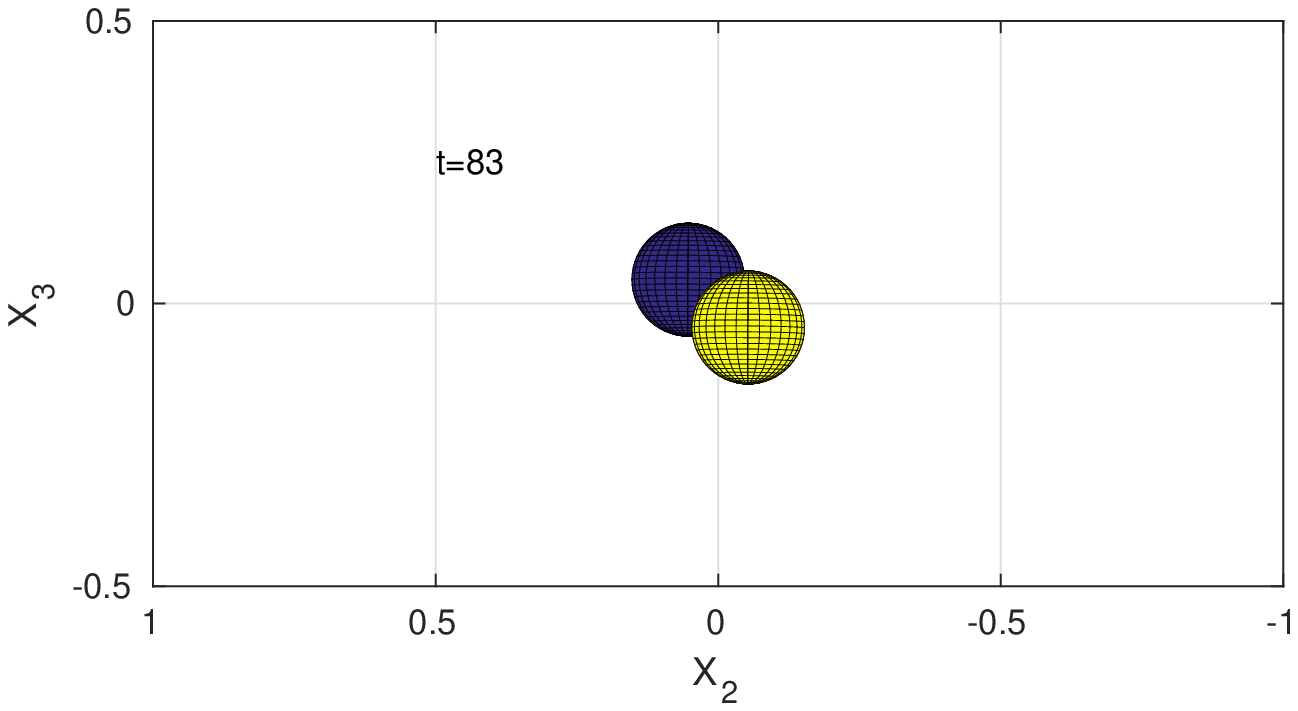}
\epsfxsize=1.95in
\epsffile{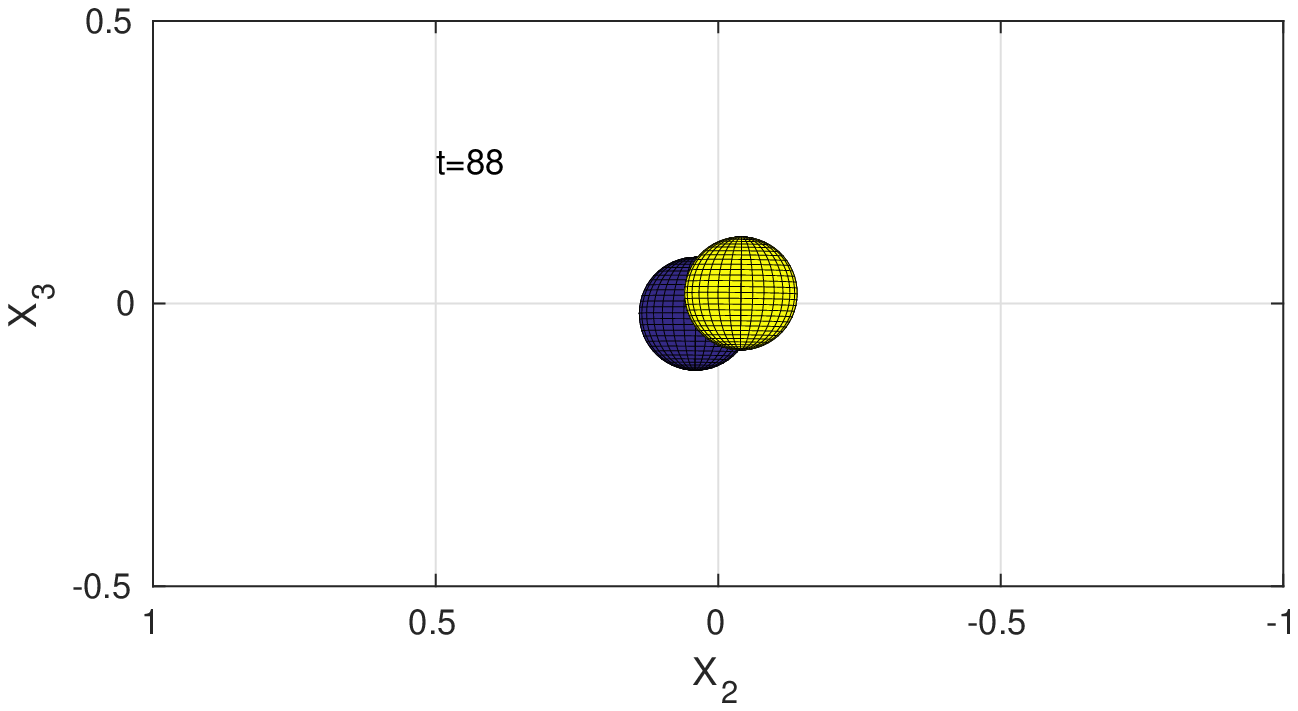}\\
\epsfxsize=1.95in
\epsffile{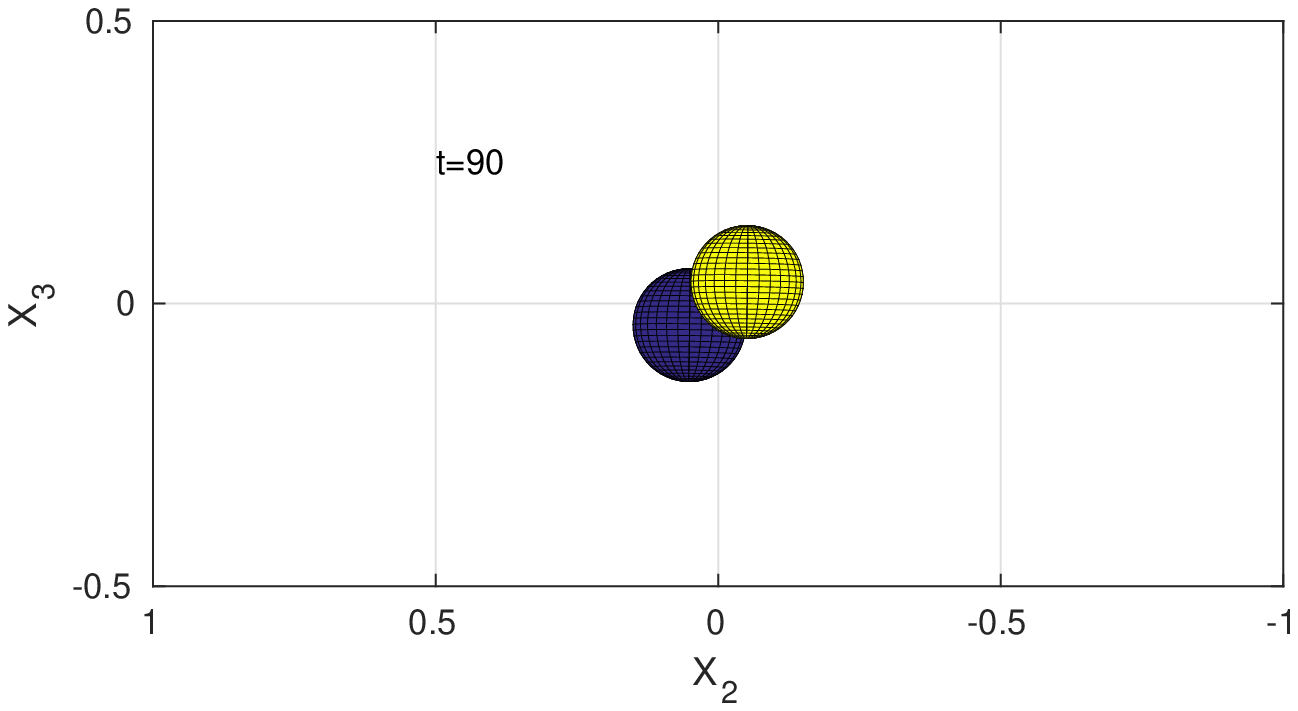}
\epsfxsize=1.95in
\epsffile{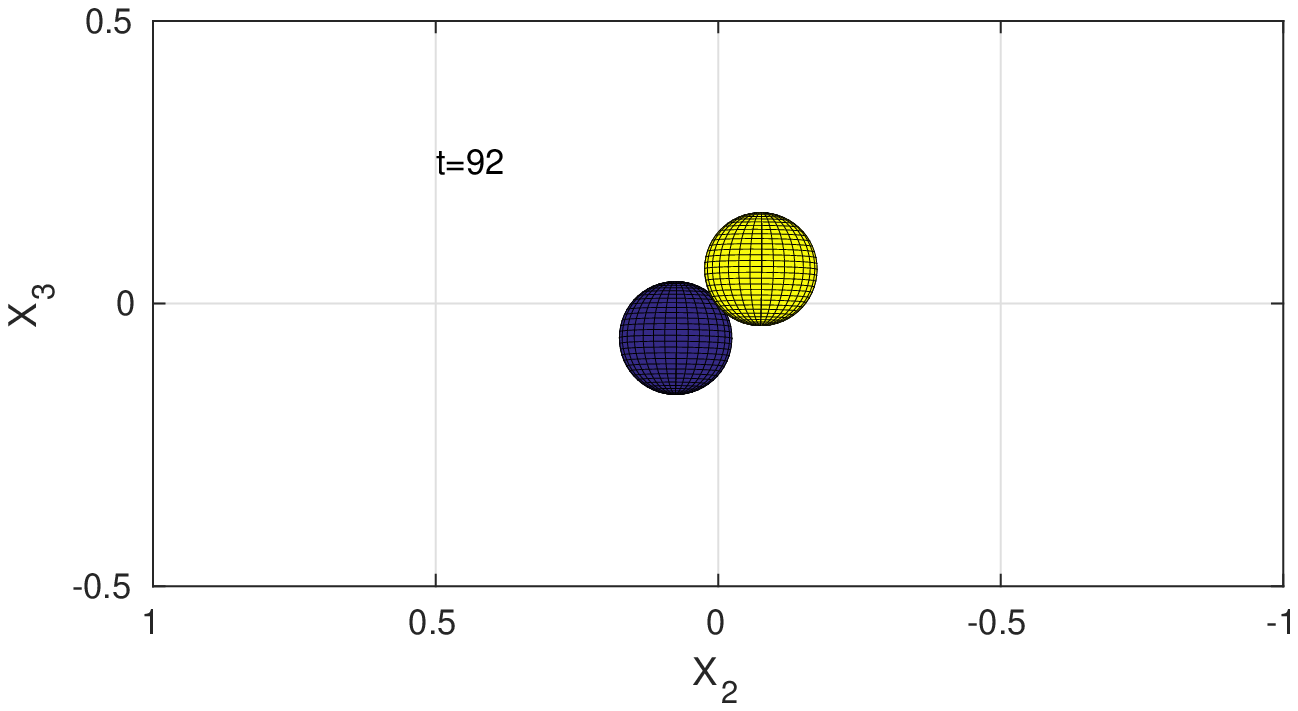}
\epsfxsize=1.95in
\epsffile{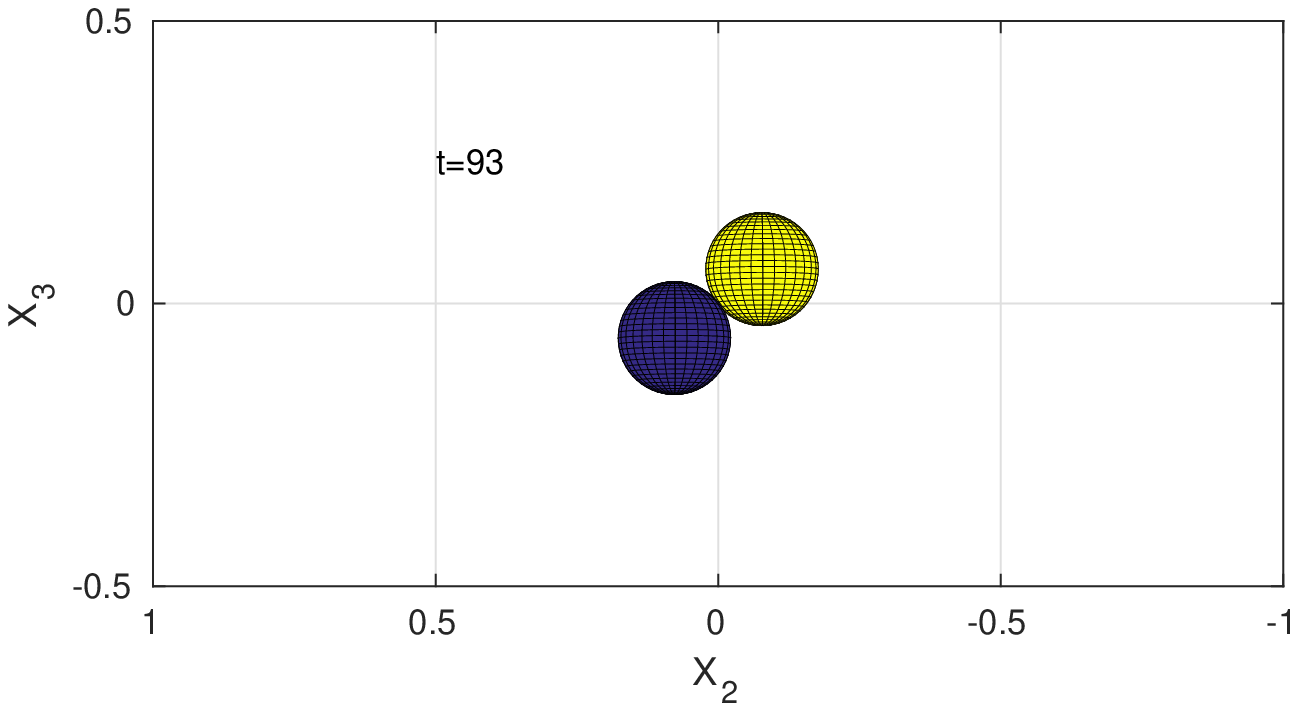}
\end{center}
\caption{The  ball position of the kayaking motion viewed in the $x_1-$direction for Wi=1 and $D=0.255$ at $t=69$, 71,  80, 81, 83, 88, 90, 92, and 93 
(from left to right and top to bottom): the kayaking motion of the two ball mass centers is rotating about the $x_2-$direction.}\label{fig:15a}
\end{figure} 

For the two balls interacting in an Oldroyd-B fluid with the same setup and initial position, we have summarized the 
results for Wi=0.1, 0.25, 0.5, 0.75, and 1 in Figs. \ref{fig:8a} to  \ref{fig:12a}.
As in Newtonian fluids, there are results of pass and return trajectories concerning two ball
encounters; but {the trajectories}  of the two ball mass centers lose the symmetry due to the
effect of elastic force arising from viscoelastic fluids. For example, the open trajectories
associated with $D=0.5$ for Wi=0.1, 0.25, 0.5, and 1 are closer to the mid-plane after {the} two
balls pass over/under each other. The elastic force is not strong enough to hold them 
together during passing over/under, but it already pulls the balls toward each other and 
then changes the shape of the trajectories. Thus the trajectories lose {their} symmetry.
For the higher values of Wi considered in this section,
there are less return trajectories; instead it is easier to obtain the chain of two balls once 
they run into each other.  Actually depending on the Weissenberg number Wi and {on} the
initial vertical displacement $\triangle s$,  two balls can form a chain  in a bounded
shear flow,  and then such chain tumbles. For example, for $D=0.316$, the two balls come {close} to
each other, form a chain and then rotate with respect to the midpoint between two mass centers (i.e., they tumble)
for Wi=0.1, 0.25, 0.5, and 1. The distance between two balls  in the $x_1-$direction becomes bigger for higher value of Wi.  
The details of the phase diagram of pass, return, and tumbling
are shown in Fig. \ref{fig:13a}. The range of the vertical distance for the passing over becomes 
bigger for higher Weissenberg numbers. For the shear {flows} considered in this article,  
{increasing}   the Wi with a fixed shear rate is {equivalent} to increase the shear 
rate with a fixed relaxation time. This explains why, for Wi=1, two balls can have {a} bigger gap 
between them while rotating with respect to the middle point between {the} two mass centers since the 
two balls are kind of moving under higher shear rate.  Those trajectories of the tumbling motion are similar to
the closed streamlines around a freely rotating ball centered at the origin shown  in Figs. \ref{fig:8a} to  \ref{fig:10a}.

For both higher values, Wi=0.75 and 1, the tumbling motion can change to  kayaking motion later on  as shown in, 
e.g., {Fig.} \ref{fig:15a} for Wi=1 and $D=0.255$.  
The chain of two balls can be viewed as a long body, even though they are not rigidly connected.
For a rigid long body rotating in a bounded shear flow of a Newtonian fluid, its stable motion is its long axis tumbling in the shear plane (i.e., the $x_1x_3-$plane) 
due to the effect of the particle inertia (e.g., see \cite{Einarsson2015}). 
Thus for the cases of lower values of Wi considered  above, i.e., Wi=0.1, 0.25 and 0.5, the tumbling motion of two balls in the shear plane is consistent with the 
stable motion of the long body in a  Newtonian fluid.  
For Wi=0.75 and 1, those numerical results of the kayaking motion suggest that
for a rigid long body rotating in a bounded shear flow of an Oldroyd-B fluid, its long axis migrates out of the shear plane. 
Those numerical results do not conflict with those of an ellipsoid in a bounded shear flow of a Giesekus fluid obtained 
by D'Avino {\it et al.} in \cite{DAvino2014} and \cite{DAvino2015b} since they did not include the effect of the particle and fluid inertia in their simulations.

\section{Conclusions}

In this article, {we discussed} a new distributed  Lagrange multiplier/fictitious domain 
method for simulating fluid-particle interaction in three-dimensional Stokes flow of Oldroyd-B fluids. 
The methodology is validated by comparing the numerical results {associated with a}
neutrally buoyant ball. For the cases of two ball encounters  under creeping flow conditions in a bounded shear 
flow for the Weissenberg number  Wi up to 1, the trajectories of the two ball mass centers are  either passing over/under or  
returning if they don't chain. If the two balls form a chain, they tumble in the shear plane for the lower {values of the} 
Weissenberg number. But for higher  {values of} Wi, they can tumble first and then kayak later. Those numerical results 
{for} two ball {chains} suggest that it is worth to further study the effect of the particle inertia on the  orientation of 
an ellipsoid in a bounded shear flow of either Oldroyd-B or Giesekus {types}.

\section*{Acknowledgments}
This work was supported by NSF (grant DMS--1418308).

\end{document}